\documentclass[journal=jacsat,manuscript=article]{achemso}
\setkeys{acs}{keywords = true}
\usepackage[dvipsnames]{xcolor}
\usepackage{xr-hyper}
\usepackage{epsfig,mathrsfs,color,latexsym,subfigure,marginnote,verbatim,relsize,mathrsfs,color,array,amsmath,amsfonts,amssymb,graphicx,pifont,hyperref,float,chngcntr}
\usepackage[version=3]{mhchem} % Formula subscripts using \ce{}
\newcommand{\be}{\begin{eqnarray}}
\newcommand{\ee}{\end{eqnarray}}
\newcommand{\lb}{\label}
\newcommand{\bl}{\boldsymbol}

\author{Susy Exists}
\affiliation[Howard]
{Department of Physics and Astronomy, Howard University, Washington DC, 20059, USA}
\alsoaffiliation[Northeastern]
{Department of Physics, Northeastern University, Boston, Massachusetts 02115, USA}
\alsoaffiliation[QSMI]{Quantum Materials and Sensing Institute, Northeastern University, Burlington, MA 01803, USA}
\altaffiliation{These authors contributed equally}

\author{Sougata Mardanya}
\affiliation[Howard]
{Department of Physics and Astronomy, Howard University, Washington DC, 20059, USA}
\altaffiliation{These authors contributed equally}
\author{Robert Markiewicz}
\affiliation[Northeastern]
{Department of Physics, Northeastern University, Boston, Massachusetts 02115, USA}
\alsoaffiliation[QSMI]{Quantum Materials and Sensing Institute, Northeastern University, Burlington, MA 01803, USA}
\altaffiliation{These authors contributed equally}

\author{Tugrul Hakioglu}
\affiliation[energy]
{Energy Institute, Istanbul Technical University, Maslak 34469, Istanbul, Turkey}
\alsoaffiliation[ITU]
{Department of Physics, Istanbul Technical University, Maslak 34469, Istanbul, Turkey}
\alsoaffiliation[Michigan]
{Department of Physics, University of Michigan, Ann Arbor, MI 48109, USA}
\alsoaffiliation[Northeastern]
{Department of Physics, Northeastern University, Boston, Massachusetts 02115, USA}
\alsoaffiliation[QSMI]{Quantum Materials and Sensing Institute, Northeastern University, Burlington, MA 01803, USA}
\altaffiliation{These authors contributed equally}

\author{Jouko Nieminen}
\affiliation[Tampare]
{Computational Physics Laboratory, Tampere University, Tampere 33014, Finland}
\alsoaffiliation[Northeastern]
{Department of Physics, Northeastern University, Boston, Massachusetts 02115, USA}
\alsoaffiliation[QSMI]{Quantum Materials and Sensing Institute, Northeastern University, Burlington, MA 01803, USA}

\author{Ville J. H\"{a}rk\"{o}nen}
\affiliation[Tampare]
{Computational Physics Laboratory, Tampere University, Tampere 33014, Finland}

\author{Cem Sanga}
\affiliation[ITU]
{Department of Physics, Istanbul Technical University, Maslak 34469, Istanbul, Turkey}

\author{Arun Bansil}
\affiliation[Northeastern]
{Department of Physics, Northeastern University, Boston, Massachusetts 02115, USA}
\alsoaffiliation[QSMI]{Quantum Materials and Sensing Institute, Northeastern University, Burlington, MA 01803, USA}

\author{Sugata Chowdhury}
\affiliation[Howard]
{Department of Physics and Astronomy, Howard University, Washington DC, 20059, USA}
\email{sugata.chowdhury@howard.edu}

\title{Giant Kohn anomaly and chiral phonons in the charge density wave phase of 1H-NbSe$_2$: impact of phonon anticrossing}


\begin{document}

\begin{abstract}
Despite extensive investigations, many aspects of charge density waves (CDWs) remain elusive, especially the relative roles of electron-phonon coupling and Fermi surface nesting as the underlying driving mechanisms responsible for the emergence of the CDW vector $\bl Q_{CDW}$. It is puzzling that even though electrons interact strongly with optical phonons in many correlated systems, the actual mode softening is of an acoustic mode. Here we consider monolayer 1H-NbSe$_2$ as an exemplar system, and through an accurate computation of the phonon self-energy, including its off-diagonal components, we provide compelling evidence that the relevant mode is a longitudinal optical phonon that softens by anti-crossing several intervening phonon bands, i.e. a Kohn ladder which has been only observed previously in high temperature superconductors.  We also show that $\bl Q_{CDW}$ is fixed by the convolution of the susceptibility and electron-phonon coupling, and that the softened phonons are circularly polarized.  
\end{abstract}


In the race to miniaturize – connecting ever smaller semiconductor devices\cite{atom-transistor} with ever thinner metallic wires\cite{wire} -- quantum effects have evolved from being a small perturbation to becoming a driving force for innovation and new devices\cite{sensing}, including quantum computers based on quantum entanglement\cite{nielsen}.  Moreover, improved density functional theories (DFTs)\cite{hse,scan} are advancing our ability to describe exotic band structures such as the linearly dispersing Dirac and Weyl bands\cite{hasan-kane,weyl-dirac}, and providing a firm basis for material-specific many-body models of quantum materials. 

Anticrossing of bands is an unusual effect, an example of the strangeness of quantum mechanics, where two bands of the same symmetry cannot cross\cite{crossing}, but must maintain a minimum gap between them\cite{degeneracy}.  Thus, as the offset parameter between two bands is continuously changed, the bands first move in k-space towards each other, but instead of crossing they appear bounce off of each other and then start moving in the opposite direction with increasing separation.  What really happened is that, while the bands were hovering near their minimal separation, they were gradually exchanging their characters, so that far from the minimal gap it looks as if they have simply crossed each other.  This is a well-known effect of the off-diagonal matrix elements of the phonon dynamic matrix. An interesting manifestation of anticrossing arises when allowed crossings lead to Dirac-like dispersions and the band continuity can only be demonstrated by passing through more than one Brillouin zone.\cite{Joh}  These effects are readily captured for electronic bands in DFT, and, as we shall show, for phononic bands in density functional perturbation theory (DFPT).  

We expect anticrossing to also play a key role in one of the most important properties of phonon bands – the driving of phase transitions via soft phonons.  The softening is often caused by electron-phonon interaction, leading to a sharp depression of the phonon band in a narrow region of the momentum space, which results in Kohn anomalies\cite{Kohn} from the peaks in the electron susceptibility.  In many materials, the strongest electron-phonon coupling is induced by optical modes involving ligand ions, which cause large variations of electron hopping parameters.  However, when an optical mode goes soft, it produces not a single Kohn anomaly but a stack of anomalies, that we refer to as a Kohn ladder, see \nameref{SM1} and Fig.~\ref{fig:1}.  The soft phonon must anticross several intermediate bands, breaking the Kohn anomaly into a ladder of several disconnected steps.  By extracting the off-diagonal components of the phonon self-energy we can separate the bare phonons at high temperatures from the dressed phonons at lower temperatures, and clearly demonstrate how the displacement pattern and the strong electron-phonon coupling are passed from band to band as the phonon softens.  Here, we analyze the monolayer form of the well-known charge-density wave material, NbSe$_2$, by revising the Quantum Espresso (QE) \cite{QE_refs1,QE_refs2} code to expose the underlying phonon anticrossing effects.  

Transition-metal dichalcogenides (TMDCs) are well-known for their layered structures in which transition-metal atoms are sandwiched between two layers of chalcogens, creating a structure that supports a wide variety of electronic interactions driven by valence electron configurations and the spatial symmetries of the unit cell ~\cite{TMDC1,TMDC2,TMDC3}. The resulting phenomena include the coexistence of charge density wave (CDW) and superconductivity (SC) in NbSe$_2$~\cite{ScCdwExp,ScCdw2}. The simultaneous presence of CDW and SC orders offers a unique platform for exploring effects of complex electron correlations in condensed matter systems.

Among the TMDCs, NbSe$_2$ is distinguished as it exhibits an unusually rich tapestry of properties in both the bulk and thin films. Bulk NbSe$_2$ hosts a CDW with a periodic lattice distortion (PLD)~\cite{FengZheng} below $T_{\text{CDW}} \approx 33$~K, along with a coexisting SC phase with $T_c \approx 7.2$~K. As the number of layers in a film of NbSe$_2$ is reduced, the CDW transition becomes stronger, and superconductivity weakens. In the 1H-monolayer limit $T_{\text{CDW}} \approx 145$~K and $T_c \approx 1.9$~K~\cite{ScCdw3,ScCdw4}. Notably, the layered architecture of TMDCs also makes them amenable to strain engineering and the construction of heterostructures~\cite{strain,hetero1,hetero2}. This adaptability enables a wide range of applications in nanoelectronics, optoelectronics, and quantum devices~\cite{Vaskivskyi2015Fast,Akinwande2017Two,Clarke2008Superconducting}, where manipulation of electronic properties at the nanoscale plays a key role.

The nature of electron-phonon coupling varies across the landscape of materials.  In some cases, electrons become localized and distort the lattice around the atomic sites, effectively enhancing the electronic mass, while in other cases the phonon modulates the bond lengths, strengthening some hoppings and weakening others\cite{polaron1,polaron2}: In polaron physics, the former are known as Holstein polarons and the latter as Peierls polarons\cite{Berciu}, with drastic differences in their properties. For Holstein polarons, with increasing electron-phonon coupling, electrons become increasingly more trapped in their polaron clouds and eventually undergo a metal-insulator transition\cite{MIT}.  In contrast, for the Peierls polarons, the bonds with strengthened hoppings can link up along certain chains, so that the polaron mass can remain small even for strong electron-phonon coupling. Holstein polarons are thus unlikely candidates for bipolaron superconductivity\cite{Mott}, but this remains a possibility for Peierls polarons. These differences should persist in the CDW physics as well.  For example, many high-frequency oxygen modes in the cuprates are of Peierls form, modulating copper-to-oxygen hopping, so that the CDW transition would be expected to involve the softening of these optical modes\cite{CuprateSoften1,CuprateSoften2,CuprateSoften3},
although this softening has been observed only rarely in experiments.

The softening of an optical phonon to zero energy leads to a form of ``quantum strangeness'' \cite{Strangeness} in the sense that the soft mode Kohn anomaly must cross all lower-lying phonon modes. Modes of different symmetries can cross without interactions, but those of the same symmetry (irreducible representation (irrep) of the dynamical matrix), interact to open an energy gap or ‘anticross’, where eigenfunctions interchange their characters. We define a ``Kohn ladder'' as a Kohn anomaly which is broken up into many segments separated by anticrossings at ascending energies.  In \nameref{SM1}, we discuss a simple toy model to demonstrate how in such a ladder the phonon displacement pattern is transferred from band to band as the LO phonon softens, so that when a CDW develops, its displacement pattern resembles that of the LO phonon. 

Such a Kohn ladder has been predicted\cite{Gutzcharge} and observed\cite{Ding} in the cuprates, where the resulting CDW is consistent with the known optical phonon symmetry of the parent compound\cite{Forgan}. In NbSe$_2$, an experimental study examined but ultimately rejected the possibility of a Kohn ladder\cite{Rez}.  However, in monolayer 1H-NbSe$_2$, which lacks inversion symmetry, high-pressure calculations report two possible CDW distortions, which are consistent with breathing-in and breathing-out distortions of an optical phonon\cite{NbSe2}. Here we re-explore this issue by examining the contribution of electron-phonon coupling to the phonon self-energy to uncover what controls the CDW $q$-vector.

\section{Results/Discussion}

\subsection{N\lowercase{b}S\lowercase{e}$_2$} 
NbSe$_2$ is a canonical metallic superconducting CDW material, with the superconducting state occurring at a temperature $T_c \ll T_{\text{CDW}}$~\cite{Wilson1975,Moncton1977}. Near $T_{\text{CDW}}$, there is a strong Kohn anomaly associated with a soft phonon and a static CDW exhibiting a $3 \times 3$ PLD in real space~\cite{EPH3,EPH4}. This corresponds to a triple CDW wavevector $\mathbf{Q}_{\text{CDW}} \approx \frac{2}{3}\overline{\Gamma M}$~\cite{Soumyanarayanan2013,EPH5}.

Unlike NbS$_2$, NbSe$_2$ does not exhibit magnetic response and it is not close to any magnetic instability in its CDW phase~\cite{magnetic,magnet2}. The electronic spin-orbit splitting at the Fermi level is negligibly small compared to the CDW energy scale~\cite{Johannes2006}. The Fermi surface of NbSe$_2$ (Fig.~\ref{fig:deconv}(a) Right) is strongly anisotropic, composed of pockets around the $\Gamma$ and $K$ points, preventing perfect nesting at the Fermi energy~\cite{Flick,FengZheng,FermiGap}. These features of the electronic structure imply that the PLD-CDW instability is far from conventional and that it results from a combination of strong electron-phonon coupling and Fermi surface topology~\cite{topo1}.

Survival of the superconducting state in the presence of such a strong CDW transition indicates that the CDW and superconducting regions in $k$-space avoid each other, confirming a strongly anisotropic Fermi surface within the CDW state \cite{new_nbse2,2new_nbse2}. This unconventional picture is distinct from the commonly observed single-acoustic-mode softening accompanied by the appearance of a uniform CDW gap around the Fermi surface. Indeed, density functional theory (DFT) based calculations show that both longitudinal acoustic and longitudinal optical phonon modes have strong coupling with electrons on the Fermi surface, and that spin-orbit effects do not play a major role.

Despite decades of study, the microscopic origin of the 3$\times$3 CDW in NbSe$_2$ remains contested: simple Fermi-surface nesting alone appears insufficient \cite{Johannes2006}; ARPES reveals two gaps and Fermi-surface ‘arcs,’ challenging a uniform-gap, nesting-only picture\cite{Borisenko2009}; other work highlights strongly momentum-dependent electron–phonon coupling and its interplay with superconductivity\cite{Kiss2007}; cross-material analysis supports EPC-driven CDWs as a viable route\cite{Calandra2009}; and broader surveys show multiple CDW classes (Peierls, excitonic, EPC-dominated), underscoring why NbSe$_2$ remains debated \cite{Plummer}.

In this study, we will focus on the $T_c \ll T_{\text{CDW}} \leq T$ regime, and employ DFT-based calculations to address two questions: 
How does the optical-acoustic phonon mode mixing relate to the Kohn anomaly? And is $\mathbf{Q}_{\text{CDW}}$ determined by the peak in the charge susceptibility or by the anisotropic electron-phonon coupling?

\begin{figure*}[ht]
\centering
\includegraphics[width=\linewidth]{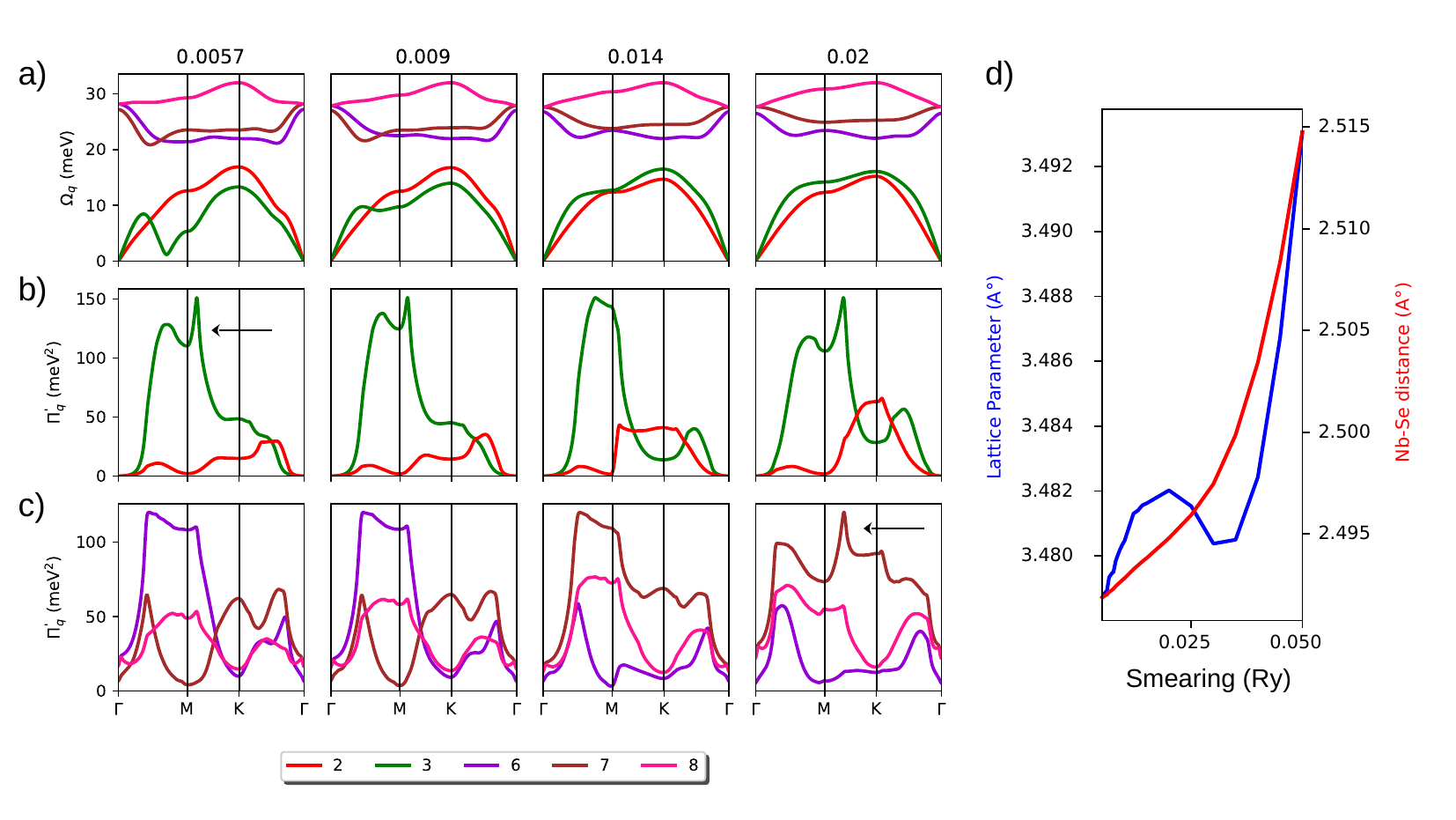}
\vskip-0.3truecm
\caption{(color online)
{\bf Phonon dispersions in NbSe$_2$}. Phonon branch indices are identified with different colors in the legend. (a) $\sigma$-dependent phonon spectrum.
(b) $\sigma$-dependent self-energy spectrum for acoustical bands. (c) $\sigma$-dependent self-energy spectrum for optical bands.
(d) Variation of unit cell parameters vs $\sigma$. Blue line gives the change in the lattice parameter, while red line gives the distance between Nb and Se atoms with respect to the change in $\sigma$.}
\label{fig:ph}
\end{figure*}

\subsection{Calculating electron-phonon coupling}
In conventional many-body approaches, the bare phonon Green's function is calculated without electron-phonon interactions, which are incorporated subsequently in the phonon self-energy~\cite{Mahan}. In contrast, DFT calculations of the phonon spectra do not make an explicit separation into bare and dressed phonons, which complicates comparisons between various phonon results ~\cite{Giustino2017,Marini2001,Heid2010,Harkonen2020}.  Here we focus on the DFT calculations, while in \nameref{SM2} we discuss differences with the many-body approach and an approximate way of relating the two approaches.

We use Quantum Espresso (QE) \cite{QE_refs1,QE_refs2} for DFT-based structural optimization and electronic structure computations, see Supporting Information (SI)-1 for details. A Fermi-Dirac type smearing $\sigma$ is invoked to simulate thermal effects and define an effective temperature $T_{\sigma}=\sigma/k_B$, where $k_B$ is Boltzmann's constant. We then use density-functional perturbation theory (DFPT), as implemented in the QE software (In the renormalization of phonons by electron-phonon coupling, Quantum Espresso does not consider a full diagonalization in the multimode phonon subspace which would require mode-mode coupling of the phonons.\cite{QuantEspRenorm} Instead QE considers the renormalization at a given phonon mode fixed in the absence of the electron-phonon interaction. In the presence of a number of phonon modes coupled to a single electron branch, the mode-mode coupling can make significant contributions to the phonon softening.), to directly calculate phonon frequencies as a function of $\sigma$. Electron-phonon coupling matrix elements are then computed by using the EPW package \cite{EPW}.  The original EPW code gives the absolute value of the electron-phonon coupling and the diagonal elements of the self-energy matrix. To assess the couplings between the phonon modes and the phonon self-energy, we modified the EPW code\cite{EPWmodified} to calculate the off-diagonal elements $\lambda\ne \lambda^\prime$ of the phonon self-energy\cite{typy} and the coupled phonon eigenstates. This mode coupling turns out to be an essential element in strengthening the CDW-PLD formation; see sections \nameref{subsec:multimode} and \nameref{section:model-kohn}.

\begin{figure*}[ht]
\centering
\includegraphics[scale=0.6]{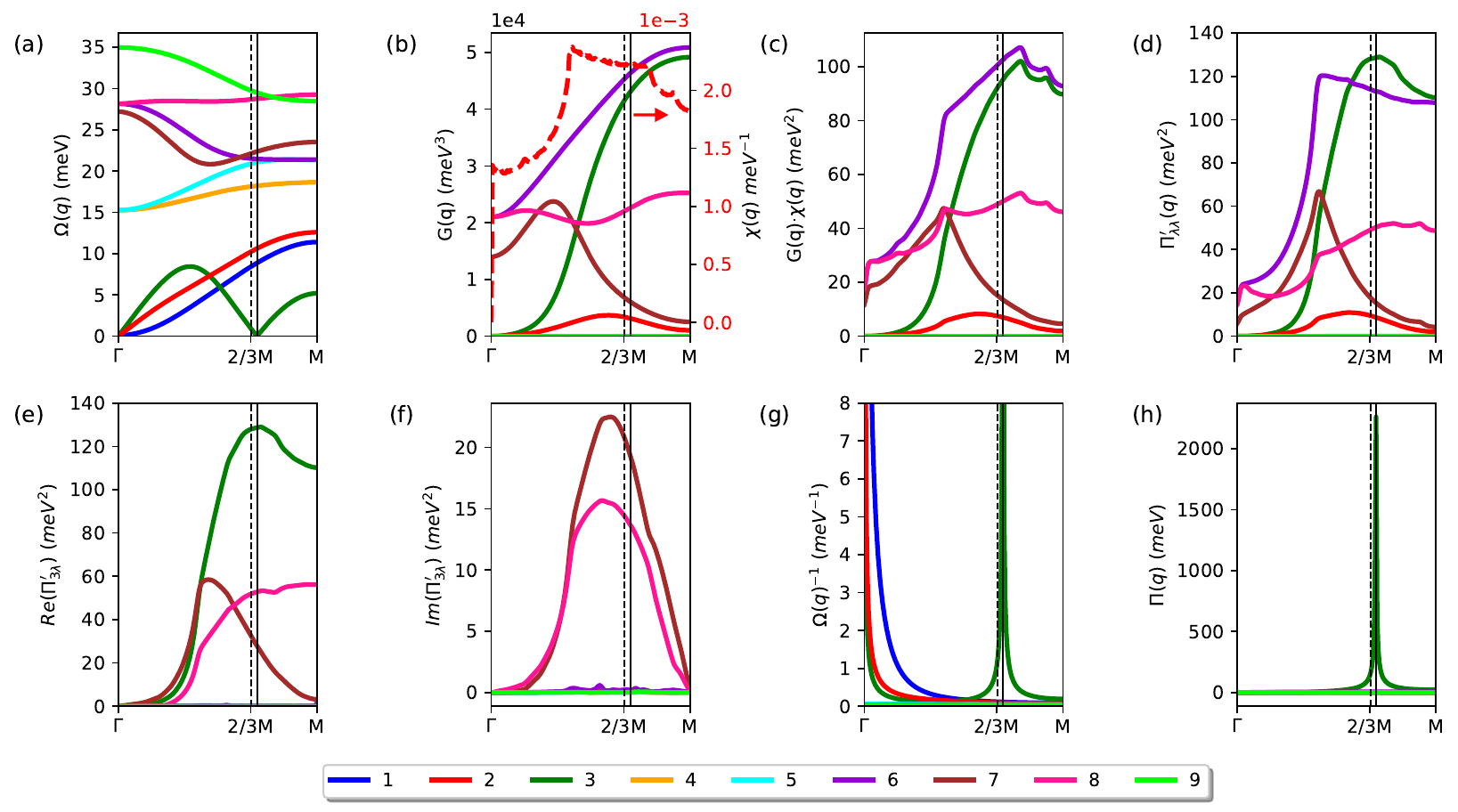}
\vskip-0.3truecm
\caption{(color online)
Decomposition of self-energy to determine the origin of the $q$-vector close to the instability temperature, $\sigma=0.00565$~Ry.  Top row (left to right): (a) Phonon dispersions for various modes, numbered with different colors in the legend. (b) Susceptibility, $\chi(q)$ (dashed red line, right-axis)), and $k$-averaged electron-phonon coupling $G(q)$ (remaining curves, left axis). (c) $G(q)\chi(q)$. (d) Diagonal self-energy $\Pi'_{\lambda \lambda}$. (e,f) Real (e) and imaginary (f) parts of the off-diagonal self-energy $\Pi'_{3\lambda}$, where 3 denotes the soft-mode phonon mode (green).  (g) $1/\Omega$. and (h) $\Pi$. Positions of the soft-mode $\bf Q_{CDW}$-vector (solid black vertical lines) and $\bf q=2/3\,\overrightarrow{\Gamma M}$ (dashed black vertical lines) are marked.} 
\label{fig:6a}
\end{figure*}

Figure~\ref{fig:ph} summarizes the evolution of the phonon spectra (a-c) and lattice parameters (d) for various $\sigma$ values. Strong $T$-dependence of the phonon frequencies and the appearance of phonon softening are clear evidence that the interactions are inseparably incorporated in the DFT calculations. Strong $T$-dependence of the atomic positions and lattice constants in the numerical results shown in Figure~\ref{fig:ph}(d) suggest the importance of multiphonon processes (noninteracting phonons do not cause thermal lattice expansion). The phonon self-energy here is calculated non-perturbatively in terms of dressed phonons, which leaves open the questions: How does one extract the bare phonon frequencies? And is the DFT based self-energy consistent with the field theory renormalization of the bare phonon frequencies?

Figure~\ref{fig:6a} presents the phonon spectrum along the $\Gamma-M$ line in panel (a), while the remaining frames provide results for various components of the calculations to gain insight into the nature and origin of the CDW instability.  Modes 3, 6, and 8 have predominantly longitudinal symmetry and interact strongly with each other.  Modes 2 and 7 are predominantly transverse, but since the modes are strictly longitudinal or transverse only along the high-symmetry lines such as $\Gamma-M$, these longitudinal and transverse modes interact away from the high-symmetry lines. These are the only modes with strong electron-phonon coupling, and therefore, we will we focus mainly on these modes. The sharp dip in the phonon dispersion of mode 3 in panel (a) is the Kohn anomaly that drives the CDW instability\cite{Kohn}.

To reconstruct the dynamical matrix in the presence of electron-phonon coupling, we start with a general framework where multiple phonon modes interact with multiple electron bands in the vicinity of the Fermi energy. [{\it Mode} and {\it band} are used here in the Born-Oppenheimer sense where the ionic and electronic degrees of freedom are decoupled.] The  electron-phonon interaction is handled perturbatively, see SI for details. 

A phonon self-energy matrix $\Pi_{\lambda \lambda^\prime}$ is introduced which includes off-diagonal coupling between the phonon modes $\lambda$ and $\lambda^\prime$ as well as the electron-phonon vertex function $g_{n,m,\lambda}(\bl k;\bl k+\bl q)$; here the electron state of wavevector $\bl k$ in band $n$ interacts with the electron state $\bl k+\bl q$ in  band $m$ by exchanging a phonon with wavevector $\bl q$ in the mode $\lambda$. Since NbSe$_2$ only has one band crossing the Fermi energy, we take $n=m=1$, and remove the electron band index for simplicity. Spin-orbit coupling, which is quite small in NbSe$_2$, is ignored.

In the multi-mode phonon basis\cite{Giustino1}, the renormalized phonons are given by the poles of the interacting Green's function:
\be
\det\Bigl\{(\Omega^2-{\bl \omega_{\bl q}^2})\operatorname{\bl I} -2\, {\bl \omega_{\bl q}} \, {\bl \Pi}(\bl q,\omega_q)\Bigr\}=0
\lb{renormalized_omega_1}
\ee
where $\bl \omega_{\bl q}$ is a diagonal matrix with elements $\omega_{\lambda,\bl q}$ in the phonon eigenmode space describing the bare-phonon frequencies and ${\bl \Pi}(\bl q,\omega)$ is the selfconsistent electron-phonon self-energy given by,
\be
\Pi_{\lambda,\lambda^\prime}(\bl q,\omega)&=&\sum_{n,m,\bl k,\sigma}\,G_{\lambda,\lambda^\prime}(n,\bl k; m,\bl k+\bl q) \nonumber \\ 
&\times& ~~~~~~\frac{f(\xi_{m,\bl k+\bl q,\sigma})-f(\xi_{n,\bl k,\sigma})}{\xi_{m,\bl k+\bl q,\sigma}-\xi_{n,\bl k,\sigma}+\omega}
\lb{ph_SE_1}
\ee
with $\xi_{n,\bl k,\sigma}$ describing the electron energy in band $n$, measured from the Fermi level and $f(x)$ is the Fermi-Dirac distribution.  Here   
\be  
G_{\lambda,\lambda^\prime}(n,\bl k;m,\bl k+\bl q)=\mathfrak{g}_{n,m,\lambda}(\bl k,\bl k+\bl q)\,\mathfrak{g}_{m,n,\lambda^\prime}(\bl k+\bl q,\bl k) 
\lb{g_square}
\ee 
where 
\be 
\mathfrak{g}_{n,m,\lambda}(\bl k;\bl k+\bl q)&=&\Bigl(\frac{\hbar}{2M\omega_{\lambda,\bl q}}\Bigr)^{1/2} g_{n,m,\lambda}(\bl k;\bl k+\bl q) ~~~~~~
\lb{micr_e-ph_1}
\ee
and $g_{n,m,\lambda}(\bl k;\bl k+\bl q)$ is the microscopic complex electron-phonon coupling given by\cite{Giustino1}  
\be 
g_{n,m,\lambda}(\bl k;\bl k+\bl q)=\langle  \Psi_{\bl k+\bl q, m} \vert  \nabla_{\bl q} V_{\bl q}.\bl \xi_{\lambda \bl q}\vert \Psi_{\bl k, n}\rangle.
\lb{e-ph_matrix_elements}
\ee 
where $\vert \Psi_{\bl k, n}\rangle$ are the electron Bloch states, see \nameref{SM2} for details. We also note that $g_{n,m,\lambda}(\bl k;\bl k+\bl q)=g_{m,n,\lambda}^*(\bl k+\bl q;\bl k)$ by  unitarity. Inhomogeneous $\bl k$ dependence in Eq.(\ref{e-ph_matrix_elements}) and hence in Eq.(\ref{g_square}) is the result of the anisotropy and the sensitivity of the Bloch wave functions to the crystal structure. 

In NbSe$_2$, the relevant sections of the Fermi surface responsible for the strong electron interactions are in the K-pockets near the zone boundaries.  In NbSe$_2$, since only one electronic band intersects the Fermi surface, we can neglect the $n,m$ indices.  The second term on the right-hand side of Eq.(\ref{ph_SE_1}) is closely related to the Lindhard electronic susceptibility (dropping the band index n and considering the spin degeneracy factor 2),
\be 
\chi_0(\bl q,\omega)=2\,\sum_{\bl k}\, \frac{f(\xi_{\bl k+\bl q})-f(\xi_{\bl k})}{\xi_{\bl k+\bl q}-\xi_{\bl k}+\omega+i\delta},
\lb{Lindhard-susceptibility} 
\ee
which we calculate using Maximally Localized Wannier Functions (MLWFs) \cite{Wannier}. We further simplify the computation by taking the adiabatic limit ($\omega \to 0$). [Non-adiabatic corrections in NbSe2 are found to be small.] Effects of the $\bl k$ dependence of $g_{n,m,\lambda}(\bl k;\bl k+\bl q)$ can be seen in $\Pi_{\lambda,\lambda^\prime}$ evaluated using the limiting cases given by Eqs. (\ref{ph_SE_1}) and (\ref{SE-to-Lindhard_susceptibility}). 
 
Note that $\Pi_{\lambda \lambda^\prime}$ depends on the bare phonon frequencies, $\omega_{\lambda \bl q}$, but that an exact solution for the dressed phonons can be obtained in principle by replacing self-consistently the bare terms by the dressed terms \cite{Hedin,Mahan}.  Below, we will refer to the replacement of $\omega_{\lambda \bl q}$ by $\Omega_{\lambda \bl q}$ as the Hedin model.

In the limit of negligibly weak lattice potential, the Bloch states can be approximated by plane waves, so that $g_{n,m,\lambda}(\bl k;\bl k+\bl q)$ in Eq.(\ref{e-ph_matrix_elements}) becomes dependent only on the momentum exchanged with the phonons\cite{Grimvall}.  The phonon self-energy in Eq.(\ref{ph_SE_1}) can then be written as:
\be 
\Pi_{\lambda,\lambda^\prime}(\bl q,\omega)&=& G_{\lambda,\lambda^\prime} (\bl q) \chi_0(\bl q,\omega)
\lb{SE-to-Lindhard_susceptibility}
\ee 
where $G_{\lambda,\lambda^\prime} (\bl q)$ matrix elements are the single mode version of Eq.(\ref{g_square}) in the plane wave approximation. Eq.(\ref{SE-to-Lindhard_susceptibility}) also represents the {\it dirty limit} phonon self-energy as discussed in Section \nameref{sec:q}.

We now discuss how the $\bl k$ dependence of $g_{n,m,\lambda}(\bl k;\bl k+\bl q)$ drives an unconventional phonon softening and CDW formation distinct from the Peierls mechanism\cite{gruner1994density}.  Figure~\ref{fig:6a}(a) shows the DFPT phonon dispersions\cite{Baroni} calculated at a temperature slightly above $T_{CDW}$. The CDW transition is accurately predicted at the wavevector $\bl Q_{CDW}\simeq 2\,{\overrightarrow{\Gamma M}}/3$ which agrees with the corresponding experimental value, see \nameref{SM2} for details. It is convenient to introduce a new quantity here:
\begin{equation}
\Pi^\prime_{\lambda \lambda^\prime}(\bl q,\omega)=\sqrt{\omega_{\lambda \bl q} \omega_{\lambda^\prime \bl q}}~\Pi_{\lambda \lambda^\prime}(\bl q,\omega),
\label{eq:7}
\end{equation}
so that Eq.(\ref{ph_SE_1}) can be recast as 
\begin{equation}\label{eq:8}
\begin{split}
&\Pi'_{\lambda,\lambda^\prime}(\bl q,\omega)=2\frac{\hbar}{2M}\sum_{n,m,\bl k}\frac{f(\xi_{m,\bl k+\bl q})-f(\xi_{n,\bl k})}{\xi_{m,\bl k+\bl q}-\xi_{n,\bl k}+\omega} \\ 
&\times~~~~~~\,g_{n,m,\lambda}(\bl k,\bl k+\bl q)\,g_{m,n,\lambda^\prime}(\bl k+\bl q,\bl k) .
\end{split}
\end{equation}

We will refer to $\Pi'$ as an auxiliary self-energy since it has dimension (energy)$^2$. Note that the overall phase $e^{i\varphi_{\bl k}}$ as a result of the gauge transformation of the Bloch eigenfunction $\vert \Psi_{\bl k}\rangle$ in Eq.(\ref{e-ph_matrix_elements}) in the context of the gauge non-invariance of $\mathfrak{g}_{\lambda}(\bl k;\bl k+\bl q)$ in the single band limit does not affect the off-diagonal elements of $\Pi'$.

To understand the origin of phonon softening, we examine {\it ab initio} and many-body results based on Eqs. (\ref{ph_SE_1}) and (\ref{SE-to-Lindhard_susceptibility}). This requires a modification of the QE+EPW results, since the DFPT already incorporates the main phonon renormalizations (\nameref{SM3}). To calculate $\Pi^\prime$ in Eq.(\ref{eq:7}), we must invert the dynamical matrix to get a handle on the nature of the bare phonons.  This involves two issues. (i) There is ambiguity in inverting the dynamical matrix, which we resolve in Sections {\nameref{section:model-kohn}} and \nameref{SM3}. And (ii) we need the off-diagonal self-energy elements, which are essential for the adiabatic term, even though these are unimportant for the non-adiabatic corrections; extraction of off-diagonal self-energies elements is discussed in \nameref{SM2}.  

\subsection{What determines $\bl Q_{CDW}$?}  
\label{sec:q}
A debate has arisen as to whether $\bl Q_{CDW}$ is dominated by the susceptibility in Eq.(\ref{Lindhard-susceptibility}) or the electron-phonon matrix elements in Eq.(\ref{g_square}). However, there is another potential mechanism. In Eq.(\ref{micr_e-ph_1}) , the pre-factor contains $1/{\omega_{\bl q}}$ in $|\mathfrak{g}|^2$. This, following Hedin\cite{Hedin}, can be replaced by $1/{\Omega_{\bl q}}$, which acts as a Stoner denominator and causes the diagonal element $\Pi_{\lambda \lambda}$ to diverge when $\Omega_{\bf q} \to 0$ for $\bl q=\bl Q_{CDW}$ as $T \to T_{CDW}$.  A similar argument has been invoked for the CDW physics in the cuprates.\cite{Gutzcharge}  

Fig.~\ref{fig:6a} presents a decomposition of the self-energy of all phonon modes in panel (a) along the $\Gamma\rightarrow M$ symmetry line. We first focus on the soft-phonon mode (dark green lines). $\Pi$ (panel (h)) shows a sharp, nearly divergent peak at the soft-mode $\bl q$-vector, which arises entirely from the $1/\Omega_q$-factor (panel (g)). In contrast, $\Pi'=\omega\Pi$ (panel (d)) has a weak peak at the same $\bl q$-vector.  This sharpening of a weak peak is a typical effect of a Stoner denominator. If the electron-phonon matrix element were $\bl q$-independent, then $\Pi'$ would peak at the same $\bl q$-value as the susceptibility $\chi$ (dashed red curve in panel (b)), which is not the case.  Similarly, if $\chi$ were constant, the peak would follow $G$ (green curve in panel (b)), but that is also not the case. The product $G\chi$ (panel (c)) comes closer, indicating that both $\chi$ and $G$ possess significant $\bl q$ and $\bl k$ dependence, such that the full integral over $\bl k$ correctly reproduces the self-energy peak in panel (d). The peaks in both $\Pi$ and $\Pi'$ (solid-black vertical lines) are in reasonable agreement with the experimental value $\bl Q_{CDW}\simeq 2~{\overrightarrow{\Gamma M}}/3$ (black vertical dashed line).  Note however that the off-diagonal terms in $\Pi'$ [panel (f)] due to the phonon mode-mode coupling can lead to a small additional shift of the $\bl Q_{CDW}$. 

When we extend the preceding discussion to encompass the self-energy in the full Brillouin zone, we see that (Fig.~\ref{fig:ph} (b)) the largest self-energy peak falls along the $M\rightarrow K$-line.  This sharp peak is associated with a weak nesting of the Fermi surface, see \nameref{SM2} for details. Insight can be gained by recalling that in many superconductors, the superconducting gap can be captured in the optical spectra only in the dirty limit where momentum conservation breaks down\cite{MatBar}. A similar mechanism may be at play with the second peak along the $M\rightarrow K$-line in our case and its absence in experiments. Note that Figure~\ref{fig:ph}(b) is for the clean limit. To capture the dirty limit, however, one must incorporate the breaking of the momentum conservation due to random impurities, which requires averaging $G_{\lambda,\lambda^\prime}$ over $\bl k$, leading to the form of Eq.(\ref{SE-to-Lindhard_susceptibility}). Fig.~\ref{fig:6hh}~(b) (\nameref{SM2}) shows that the averaging procedure smears out this weak nesting feature while preserving the peak along $\Gamma-M$ with a modest loss of intensity.  A similar conclusion was reached in Ref.~\cite{Plummer}.

\subsection{Coupling of bare phonon bands}  To extract the bare phonons, the divergent $1/\Omega_{\lambda,\bl q}$-factor in Fig.~\ref{fig:6a}(g) must be replaced by  $1/\omega_{\lambda,\bl q}$ (SM 3).  Once this is done, we can go beyond the above single mode picture by analyzing the $T$-dependence of the diagonal self-energies in Figs.~\ref{fig:ph} (b) and (c). We discuss the resulting anticrossing phenomena by considering two distinct cases below.

\subsubsection{Longitudinal-transverse anticrossing}
Figure~\ref{fig:ph}~(a) shows that reducing the temperature (right to left) results in crossing of the modes 2 and 3 and of modes 6 and 7 along the $\Gamma-M$ line, see Fig. \ref{fig:6a} for notation of phonon modes. Note that the optical modes 6 and 7 display at low temperatures an unusual Mobius topology along the $\Gamma \to M \to K \to \Gamma$ line: if one follows the dispersion of mode 6, starting from the left-hand side $\Gamma$ point, one ends up on mode 7 on the right-hand side.  Repeating this process into the next $\Gamma$-point by using symmetry to jump to the left-hand side $\Gamma$-point, and starting on mode 7, we see that only after winding twice about this path do we return to the original point. This behavior can be understood by noting that the longitudinal and transverse modes are orthogonal, so that we normally expect them to cross. However, in an anisotropic material, eigenmodes are strictly longitudinal/transverse only along the high symmetry lines but this is not the case along a general direction in the Brillouin zone, so that the bands then must anti-cross and exchange wavefunction character. Since modes 6 and 7  only cross once along $\Gamma-M$, they must wind twice to go from $6\rightarrow 7\rightarrow 6$ before returning to the starting mode.

A similar phenomenon is at play for modes 2 and 3 in Fig.\ref{fig:ph}, although it is hard to see the crossing between these modes because they are acoustic modes with vanishing frequencies at $\Gamma$. However, the most dispersive mode near $\Gamma$ is longitudinal (L, numbered 3) at the left-hand side $\Gamma$-point. The  resulting mode is numbered 2 after the $\Gamma \to M \to K \to \Gamma$ loop is closed. Similar winding numbers have been observed in the cuprates under pressure\cite{Joh}, see also Ref.~\cite{Zak}. At higher temperatures (right column in Fig.\ref{fig:ph}) both anti-crossings disappear, suggesting that they are  related to phonon softening and electron-phonon interactions. Notably, crossing of bands 6 and 7 is symmetry protected only along the $\Gamma M$ line, where the crossing point forms the tip of a phononic Dirac dispersion. Since this band crossing is driven by phonon softening, it constitutes an electron-phonon-coupling induced topological phase transition.

Insight into the crossing of modes 6 and 7 is also obtained by looking at the self-energies. As $T$ increases, the longitudinal mode 7 hardens and stops overlapping the transverse mode. At the crossing point, a part of the self-energy, especially along $K$ to $\Gamma$ jumps from the lower to the higher (now longitudinal) mode. [The strong spectral weight is always associated with the upper modes near $K$, but due to the change of connectivity, the labels of the modes change.]  Effects of anticrossing can, however, still be seen in Figs.~\ref{fig:ph}(b,c), where the spectral weight of the upper mode decreases substantially below the anticrossing temperature, i.e., as $T$ is lowered, the modes start to anticross, and more longitudinal character transfers to the lower modes, {\it along with the strong electron-phonon coupling}.

Turning to modes 2 and 3, the crossing along $\Gamma M$ is seen to disappear at essentially the same temperature as for modes 6 and 7, and there is also a transfer of self-energy along $K- \Gamma$, but the effect is less pronounced because the modes do not actually cross.   

\subsubsection{Longitudinal-longitudinal anticrossing}
Here, the anticrossing of the three longitudinal modes is harder to see in the dispersions as there are no symmetry-allowed crossings. Kohn-anomaly related anticrossings may thus be taken as a demanding test of a phonon code. Although we have shown that QE captures mode-crossing effects for modes 6 and 7, let us now consider modes 3 and 8, which are the lowest and highest longitudinal modes.  As expected, the self-energy of mode 3 decreases with increasing $T$.  However, the self-energy of mode 8 increases with increasing $T$, and as Figs.~\ref{fig:ph}(b,c) indicate, this is due to transfer of spectral weight from mode 3 (note the anomalous peak along the $M K$ line marked by the arrow in panel (b)), which shifts mainly to the transverse mode 7 (arrow in panel (c)), and is starting to appear in mode 8 at the highest temperature studied. These results unambiguously demonstrate that the mode softening starts in the highest longitudinal mode, and as temperature decreases, the spectral weight transfers to successively lower longitudinal modes via mode coupling,  with further spectral weight transfer between nominally longitudinal and transverse modes, since they are not strictly orthogonal at arbitrary $\bl k$-points. In the remainder of this paper, we will explore the dynamical matrix in order to extract a coherent picture in terms of `bare' phonon frequencies.

\begin{figure}[h] 
\includegraphics[width=\linewidth]{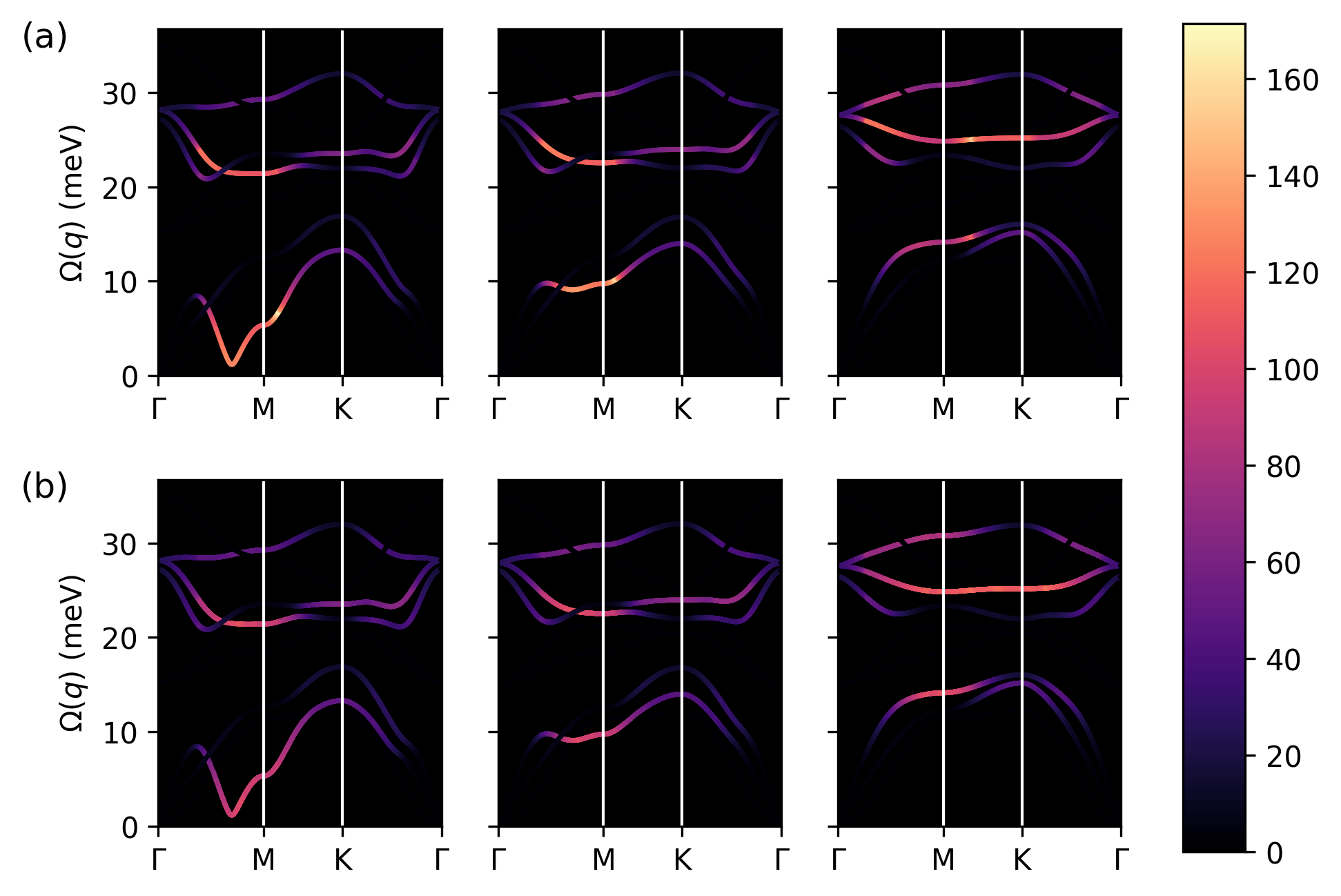}   
\vskip-0.3truecm
\caption{(color online)
{\bf Diagonal elements of phonon self-energy.} Smearing of (from left to right) 0.0057, 0.009, and 0.02 Ry is used. From top to bottom: (a) clean-limit and (b) dirty limit; projected in color on phonon bands (colors from dark to light denote weak to strong phonon self-energies in meV).}
\label{fig:7}
\end{figure}

Figure~\ref{fig:7} provides further insight by plotting the diagonal self-energy weight on top of the phonon dispersion data from Fig.~\ref{fig:ph}, with results for the clean limit given in the top row (a) and the dirty limit in the bottom row (b). Transfer of spectral weight from high to low frequencies as temperature decreases can be seen, which is somewhat clearer in the dirty limit (bottom row).

\subsection{Dynamical matrix and bare phonon frequencies}
\label{DMB}
We can obtain the bare frequencies $\omega_{\bf q}(T)$ at various temperatures by inverting the dynamical matrix. We would expect the bare frequencies to be approximately independent of $T$, but that is not the case, see \nameref{SM3} for details. Note that phonon anharmonicity\cite{anharm} is not implicit in the DFPT calculations used in this work but this can be accessed using stochastic self-consistent harmonic approximation\cite{SSCHA} implemented in a more general DFPT formulation \cite{PhysRevB.39.13120,PhysRevB.55.10355}.  
 
\begin{figure}[h]  
\includegraphics[scale=0.6]{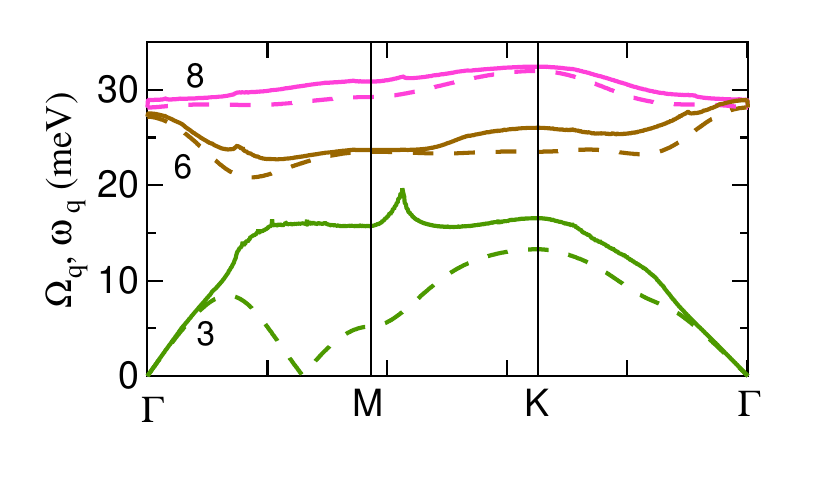}
\vskip-0.3truecm
\caption{(color online)
{\bf Single mode calculations for the three longitudinal modes (modes 3, 6, and 8)}. Solution of Eq.(SI.4) for $\sigma$ =0.0565~Ry, comparing the dressed ($\Omega_q$, dashed lines) and bare frequencies ($\omega_q$, solid lines). }
\label{fig:2b}
\end{figure}

We must first settle on the form of the dynamical matrix. In SI-1, we show that, despite some ambiguity, the appropriate choice is:
\be
\det\Bigl\{(\omega^2-{\bl \Omega_{\bl q}^2})\operatorname{\bl I} +2\, {\bl \Omega_{\bl q}} \, {\bl \Pi}(\bl q,\Omega_q)\Bigr\}=0,
\lb{renormalized_omega_3}
\ee
which provides a conventional matrix equation for the bare phonon frequencies in terms of the dressed ones. Fig.~\ref{fig:2b} shows the bare frequencies (solid lines) obtained by neglecting the off-diagonal terms in Eq.(\ref{renormalized_omega_3}).  Although phonon softening can be seen, the mode anticrossings, which are controlled by the off-diagonal terms are absent.

To understand why $\Omega_{\textbf{q}}$ becomes soft, we rewrite the single mode solution of Eq.(\ref{renormalized_omega_1}) as $\Omega_{\bf {q}}^2=\omega_{\bf {q}}^2(1-S_{\bf {q}})$, where $S_{\bf{q}}=-2\Pi'(\bf{q},\Omega_{\bf{q}})/{\omega_{\textbf{q}}^{\text{2}}}$ is an effective Stoner factor\cite{diSalvo1977}. The instability (soft phonon) occurs at that $\bf q$ for which $S_{\bf q}\rightarrow 1$ first. At $\bf Q_{CDW}$, the bare frequency is $\omega_{\bf Q_{CDW}}=\sqrt{-2\Pi'(\bf Q_{CDW})}$.  If $\omega_{\bf q}$ depends weakly on $\bf q$, then the soft mode occurs when $\vert \Pi'(\bf q) \vert$ is at its maximum.

\subsubsection{Multimode calculations}  
\label{subsec:multimode}
\begin{figure}[h] 
\includegraphics[width=\linewidth]{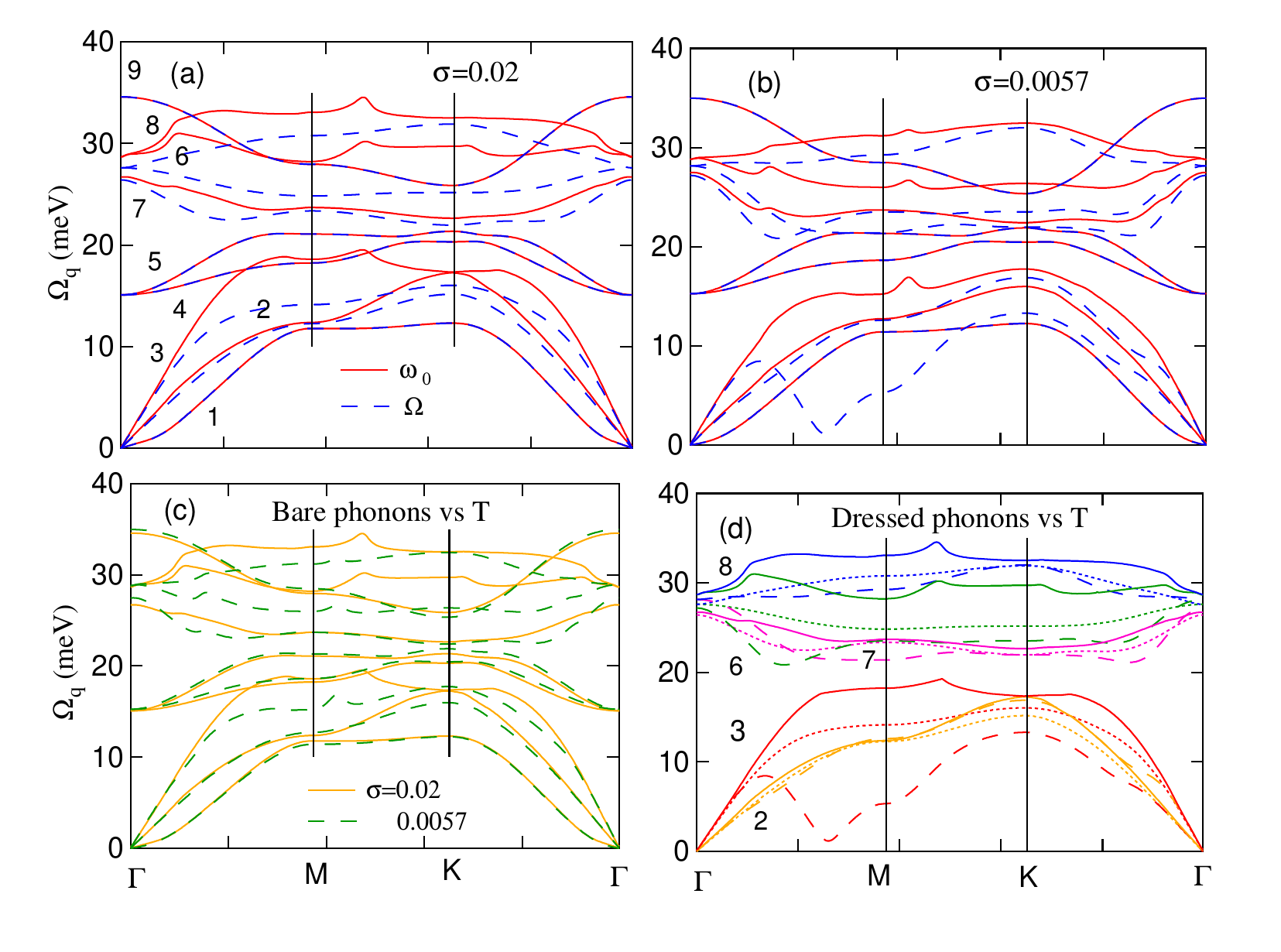} 
\vskip-0.3truecm
\caption{(color online)
(a) Dressed (blue dashed lines) and bare (red solid lines) phonons for $\sigma$ = 0.02~Ry smearing. (b) Same as (a) but for $\sigma$ = 0.0057~Ry. (c) Temperature dependence of the bare phonons for $\sigma$ = 0.02~Ry (orange solid lines) and 0.0057~Ry (green dashed lines). (d) Temperature dependence of the dressed phonons, with colors representing various phonon modes: 2 (orange), 3 (red), 6 (magenta), 7 (green), and 8 (blue). $\sigma$ = 0.0057~Ry (dashed lines) and 0.02~eV (dotted lines), while the solid lines ($\sigma$ = 0.02~Ry bare phonons) are our best estimate of the high-$T$ limit.} 
\label{fig:6b}
\end{figure}

Figs.~\ref{fig:6b}(a) and (b) compare the dressed and bare phonon modes at two temperatures ($\sigma$ = 0.02 and 0.0057~Ry) obtained by solving the full $9\times 9$ dynamical matrix. Modes 1, 4, 5, and 9 have no electron-phonon coupling, so that $\omega_{\bf q}=\Omega_{\bf q}$ at both temperatures, although  these modes still exhibit a weak $T$-dependence, panel (c). Other modes display stronger $T$-dependence, especially the longitudinal modes. We will see below that most of the $T$ dependence of the transverse modes is associated with anticrossing related weight transfer from longitudinal to transverse modes.  

Figure~\ref{fig:6b}(d) provides evidence of the Kohn ladder related to the softening of modes 8, 7, and 3.  Notably, mode 7 shows strong softening at higher $T$s, which is not as strong when it interacts with mode 6, where mode 3 starts to soften more rapidly.  Transverse modes have a weaker $T$-dependence, mainly in the $q$-regions where the self-energy shows transfer of spectral weight from longitudinal bands, and both bands show regions of phonon hardening at low temperatures.  

\subsection{A Model of a Kohn ladder}
\label{section:model-kohn}
For gaining a handle on Kohn-ladder formation, we consider a 3-band (bands 2, 6, and 8) model calculation starting at high temperature.  Following the toy model in \nameref{SM1}, we assume that all the self-energy is initially concentrated on the highest mode.  Specifically, for the model self-energy $\Pi'_m$, we assume off-diagonal elements $|\Pi'_{m,ij}|^2=20$ ($i\ne j$), $\Pi'_{m,33}=\Pi'_{m,66}$=0, and the temperature dependent self-energy is a scaled version of $\Pi'_{33}$ in Fig.~2(b), $\Pi'_{m,88}=\alpha\Pi'_{33}$, where $\alpha$ increases as $T$ decreases. 

\begin{figure}[h]   
\includegraphics[scale=0.5]{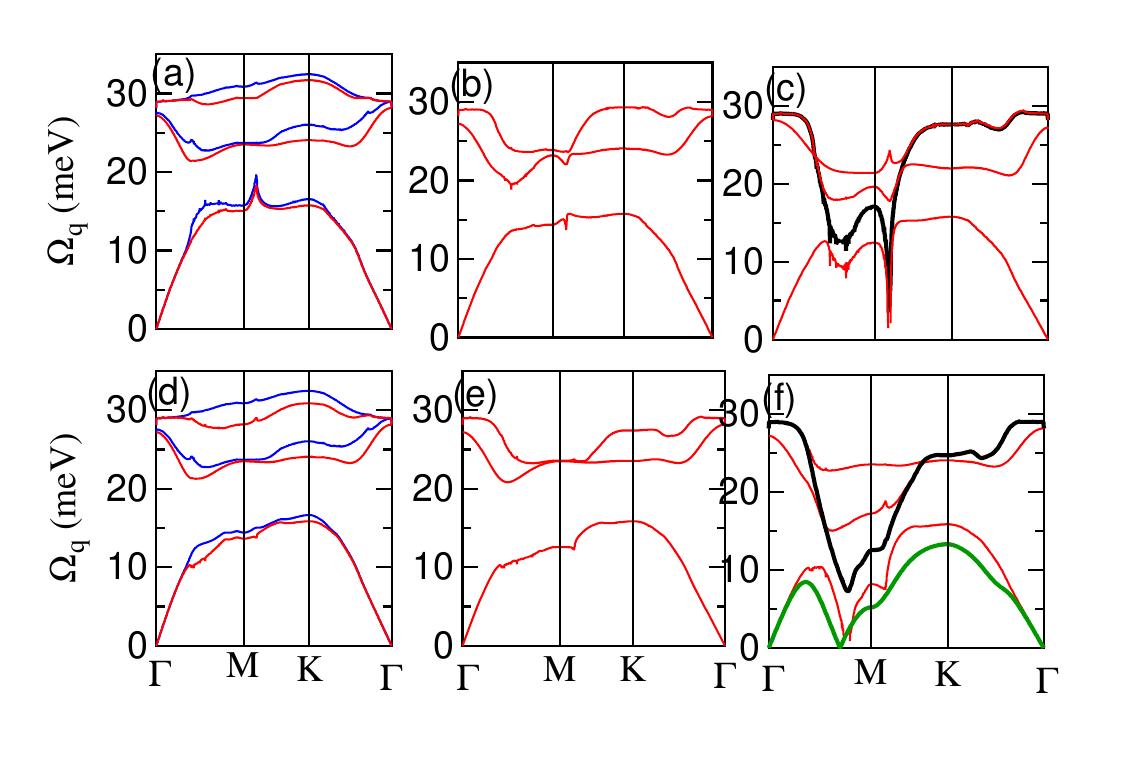}
\vskip-0.3truecm
\caption{(color online)
\textbf{Model calculations of the evolution of dressed frequencies at a series of temperatures scaled by $\alpha$}. (a-c) and (d-f) refer to the clean and dirty limits, respectively. For $\alpha=0$ (highest $T$), the one-band solutions for $\omega_q$ were calculated using $\sigma=0.03$~Ry (blue lines in (a), (d)). Lower temperatures use full $\Pi'_m$ matrix.  Red curves correspond to $\alpha$ =1 (a), 5 (b), 9 (c), 2.5 (d), 9 (e), and 12 (f).  Other curves: green line in (f) = band 3 from $\sigma=0.07$~Ry data; black lines in (c),(f) = band 8 for same $\alpha$ as red lines for $\alpha=9$ without off-diagonal elements, $\Pi'_{m,ij}=0$. 
}
\label{fig:3}
\end{figure}

Figure~\ref{fig:3} presents illustrative results, which are based on taking the bare phonons from the one-band model solutions for  $\sigma=0.00565$~Ry (blue lines in frames (a) and (d)). The results, which illustrate the anticrossing phenomena, are in general accord with the QE results (green line in (f)).  Figures~\ref{fig:3}(a-c) use the clean-limit $\Pi'_{33}$, showing that the anomalous nesting peak spoils the agreement with experiment, as expected, while the dirty limit results, Figs.~\ref{fig:3}(d-f), restore the good agreement. Figs.~\ref{fig:3}(d-f) also show that for $\alpha\le 5$, most of the softening is in the LO modes 6 and 8, while the LA mode barely changes. For larger $\alpha$, most of the softening is in the LA mode.  The latter regime resembles the QE data for $\sigma = 0.03$~Ry (green line in (f)).  

Notably, if we turn off mode-coupling, the softening of mode 8 at largest $\alpha$ (black curves in (c),(f)) is too weak for a CDW instability. This can be understood as follows. Considering two modes only, anti-crossing results in the renormalized frequencies given by: 
\be 
\Omega^2_{\pm}=\frac{1}{2}\,\Bigl[\Omega_1^2+\Omega_2^2 \pm \sqrt{(\Omega_1^2-\Omega_2^2)^2+16\omega_1 \omega_2 \vert \Pi_{12}\vert^2} \Bigr]
\lb{mode-mode-coupling-threshold}
\ee 
where the $\Omega_{1,2}$ are the renormalized frequencies found from Eqs. ~\ref{renormalized_omega_1} (without mode mixing) and $\Omega_\pm$ (including mode mixing. Clearly, $\Omega^2_-$ can go soft completely at the critical mode mixing threshold $\Pi_{12}^{(c)}$ where 
\be 
\vert \Pi_{12}^{(c)}\vert^2=\frac{\Omega_1^2\,\Omega_2^2}{4\omega_1\,\omega_2}
\lb{mode-mode-mixing-new-threshold}
\ee 
which takes place before $\Omega_1^2$ goes soft (without mode-mode coupling). Off-diagonal self-energies thus generally enhance $T_{CDW}$. Mode-mode mixing can also induce a shift in $\bl Q_{CDW}$. These effects can be shown for NbSe$_2$ by investigating the difference between $\Omega_\pm$ and $\Omega_{1,2}$.
Note that while the diagonal self-energies of Eq. (1) are real, the off-diagonal terms can be complex.  In the irreducible representation of dimension 3, the Hamiltonian determinant contains the term $-16\omega_1\omega_2\omega_3Re(\Pi_{12}\Pi_{23}\Pi_{31})$.  That is, the phonon energies depend on the relative phases between the off-diagonal components of the self-energy matrix. In contrast, for irreducible representations of lower dimension, the off-diagonal term only enters as $|\Pi_{12}|^2$.

Interestingly, the initial electron-phonon interaction at high temperatures is longitudinal, so that the main phonon softening predominantly affects longitudinal modes.  However, due to anticrossing, the phonon that softens will appear to be from a transverse mode, unless, as in NbSe$_2$, the softening is along a high-symmetry line or at a point where longitudinal and transverse modes are strictly orthogonal. 

\subsection{Discussion}

\subsubsection{Giant Kohn anomalies}

We expect strong electron-phonon coupling to often involve optical phonons which strongly modulate electron hopping integrals to yield Kohn anomalies.  The resulting Kohn ladder is captured by DFPT, including the resulting CDW whose displacement pattern matches that of the involved optical phonon.  (In simulating higher temperatures via $\sigma$, the largest $\sigma$ studied was not large enough to properly restore the large electron-phonon coupling to the highest LO branch, thereby obscuring the origin of the Kohn anomaly.)

The approach we discussed in this paper should find wide applicability for understanding anticrossing effects and the resulting giant Kohn-ladder anomalies in correlated $d-$ and $f-$electron materials. Not including effects of anticrossing could underestimate the strength of electron-phonon coupling. Even in graphene, DFT calculations fail to properly capture the Kohn anomaly\cite{graphene}. Notably, a study of spin-lattice relaxation argues that most phonons play no role in relaxation, which is dominated by phonons with rotational or intramolecular vibrational patterns.\cite{SLR1} Moreover, even the latter phonons lie at high energies, and only the small tail of these phonons at low energies dominates the relaxation process.  Our study suggests that phonons involved in relaxation are all optical, and that the low energy tail is produced by strong electron-phonon interaction in the form of Kohn anomalies, which push some modes to lower energies (soft modes). A chiral Kohn anomaly has been reported in a Weyl semimetal.\cite{Mingda}  

An alternative approach that is not based on using off-diagonal self-energies for extracting bare phonon frequencies\cite{bare} has been employed
to study phonon softening in 1H-TaS$_2$\cite{TaS2}.  Notably, Fig.~2(b) of Ref.~\cite{TaS2} shows clear evidence of a Kohn ladder remarkably similar to our findings, although the authors state: ``The bare phonon dispersion of 1H-TaS$_2$ ..., which excludes screening intrinsic to the active subspace, ... does not show any Kohn anomalies", and they thus rule out any role of phonons from other bands.  But, the analysis of Ref.~\cite{TaS2} overlooks the possibility that phonons in the LA band at low $T$ started out as LO phonons at high $T$ and passed downward via anticrossing, which is the essence of the Kohn ladder effect.

Finally, we note that it was recently reported that the off-diagonal phonon self-energy can also induce anomalous behavior in the electronic dispersion, specifically, a finite-temperature topological-to-trivial transition.\cite{ToptoNot}  Then, at the critical temperature, an allowed band crossing appears in the electronic dispersion, similar to the ones we discussed above in the phonon spectra.

\begin{figure*}[ht]
\centering
    \includegraphics[width=\linewidth]{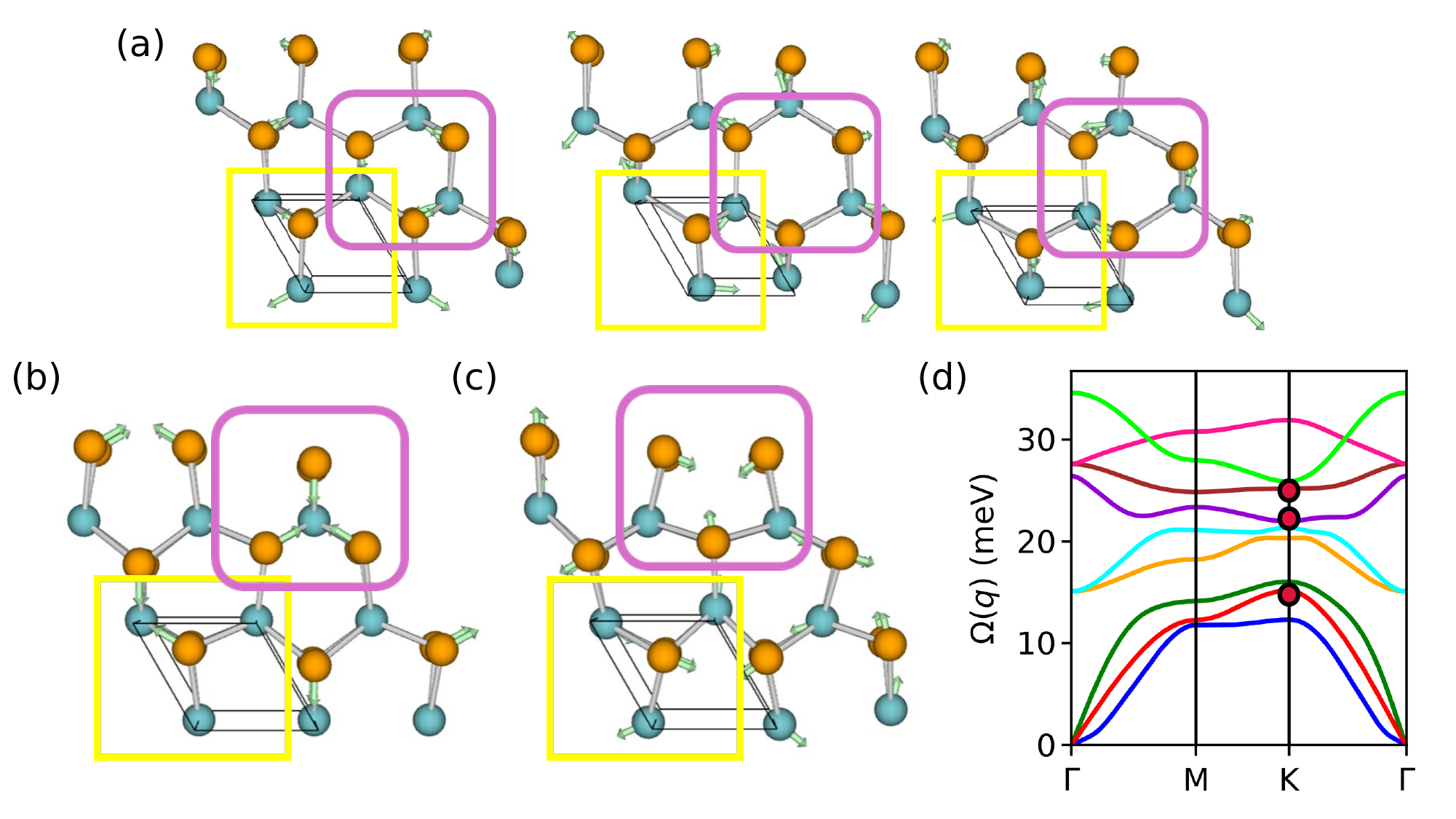}  
\vskip-0.3truecm
\caption{(color online)
{\bf K-point phonon displacements.} (a) Mode 6, showing three phases of displacement in one cycle; (b) mode 7, and (c) mode 2 showing a single phase; and (d) phonon dispersions with red dots showing $k$-points used (from top to bottom): mode 6 (a), 7 (b), and 2(c).  Yellow boxes highlight one triangle, while violet boxes highlight one hexagon.}
\label{fig:8}
\end{figure*}

\subsubsection{Visualizing soft phonons}
By using the phonon visualizer software \cite{phonon_visualizer}, we find that the phonon modes in NbSe$_2$ with significant electron-phonon couplings undergo displacements that are circularly or elliptically polarized, raising the interesting question as to what happens when a circular phonon softens. For a linearly polarized phonon, the frozen phonon CDW will have the maximum distortion associated with the soft phonon. But circular phonons have both amplitude and phase as variables, and the phase variable could lock in uniformly, adopt spatially random fixed directions (as in order-disorder transitions\cite{OD}), or remain free as in a time crystal\cite{time1,time2}. There is also a striking similarity between the circular phonon displacements and the dynamic Jahn-Teller (dJT) phenomena \cite{Bersuker}, which have been suggested as a source for time crystals.  Importantly, many correlated materials are characterized by poorly understood pseudogaps, where ``intertwined orders" prevent the establishment of any long-range order.  In such a case, a phonon would soften only to a low, but finite frequency, leaving a short-range order with strong fluctuations.  While this does not appear to be the case in NbSe$_2$, there are many materials of interest, including cuprates and nickelates, where pseudogap phenomena are observed.  If the soft phonons are circular or chiral, the resulting fluctuations would bear a strong resemblance to time crystals, without violating any of the no-go theorems for a time-crystal ground state.
It is interesting to consider two $q$-values here: $\bf q=\overrightarrow{\Gamma K}$, which lies in the region where the longitudinal-transverse anticrossing occurs and $\bf q=2/3 \overrightarrow{\Gamma M}$, where the soft phonon lies.  These two $q$ points are special in that they are 1/3 of a combination of reciprocal lattice vectors.  Remarkably, most of the $K$-point phonons are associated with recognizable dJT displacements.  One of the earliest dJT systems involves triangular molecules, such as Na$_3$, which can have a Berry phase that leads to an extra orbital angular momentum with half-integer quantization (a `time molecule').\cite{LongHi,Na3}  Similar dJT phenomena can be found on impurity sites in other crystals.

A layer of NbSe$_2$ can be viewed as a lattice of centered triangular molecules, either triangles of Se$_2$ molecules with a central Nb or of Nb molecules with a central Se$_2$.  For several $K$-point circular phonons, one cycle of rotation amounts to the central atom displacing sequentially toward each of the triangular atoms, leading to a time-averaged symmetrical pattern (yellow squares in Fig.~\ref{fig:8}), as in Refs.~\cite{LongHi,Na3}.  However, a lattice of dJT molecules has extra degrees of freedom, which also have a dJT interpretation.  On each of the three steps in a cycle, a pattern forms on one set of atoms, then dissolves and reforms on a second and then a third set before it repeats in the next cycle.  We have found three such patterns, enclosed by purple squares in Fig.~\ref{fig:8}: (i) `breathing in' (panel (b)), where three Se$_2$s move close to a central Nb; (ii) `breathing out' (panel (c)), where the three Se$_2$s move close to an empty site; and (iii) Kekul\'e-distortion of a hexagon, where panel (a) illustrates the three phases of a cycle.  

In connection with circular phonons, it is interesting to recall that one would intuitively expect a benzene ring (C$_6$ molecule) to have alternating single and double bonds, where the two carbons joined by a double bond are slightly closer to each other.  However, experimentally all the carbons in the benzene ring seem to be equivalent.  Kekul\'e interpreted this via a dynamic model of benzene, where the long and short bonds interchange dynamically in cycles of oscillations.\cite{Pauling}  This idea was formalized in Pauling's resonating valence bond (RVB) of benzene, which can be considered a form of dJT effect\cite{Pauling}. In our case, we have a circular phonon version of this effect.

Figure~\ref{fig:8} illustrates displacement patterns for several $K$-point phonons. Panel (a) shows the Kekulé-like pattern in a hexagon of alternating Nb and Se$_2$ pairs (violet box) in the three steps involved in one cycle of mode 6. (This is not strictly a Kekulé pattern, as a cycle contains three phases, with alternating short bonds 12-34-56, 23-45-61, and an undistorted hexagon.  To confirm our identification of this pattern, we show in SI-2 that a similar pattern arises in graphene, which is a lattice of benzene rings.) Panel (b) shows the breathing-in mode in mode 7, while panel (c) depicts the breathing-out mode of mode 2. The Kekulé state is involved in the softening between modes 6 and 7, but it is not the ground state. The true soft mode lies near 2/3 of $\Gamma-M$. At $M$, all modes are linearly polarized, but as one goes toward $\Gamma$, one set of atoms moves in the orthogonal direction to form an elliptical orbit. Notably, in 1H-NbSe$_2$, an earlier work found two static CDW ground states, corresponding to the breathing-in and breathing-out patterns, with an energy difference of $\sim$8~meV.\cite{NbSe2}

\begin{figure*}[ht] 
\includegraphics[width=0.7\linewidth]{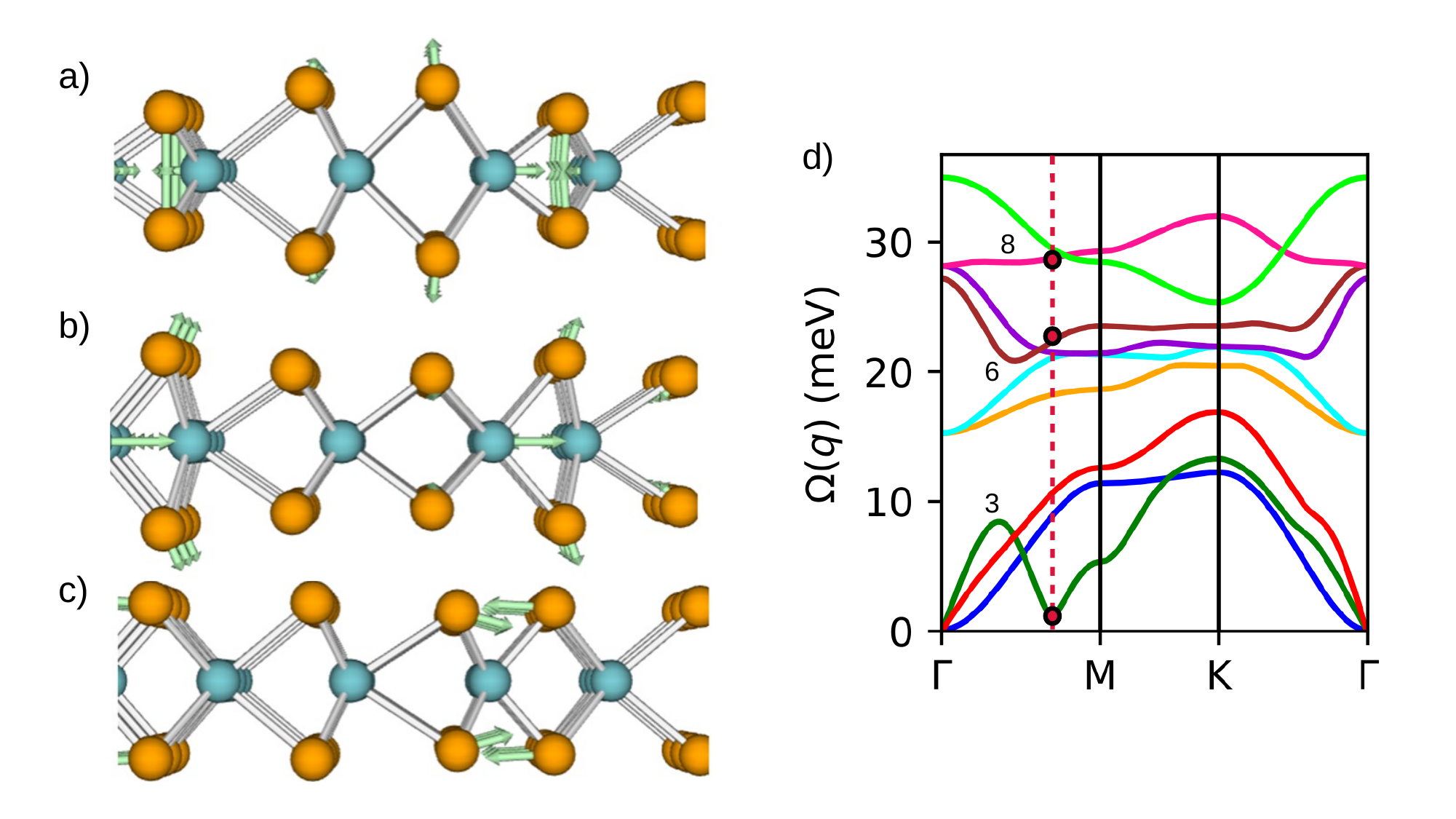}   
\vskip-0.3truecm
\caption{(color online)
{\bf 2($\Gamma$-M)/3-point phonon displacements.}  (a) to (c) show modes 8, 6 and 3, respectively, in the Y-Z plane, where ${\bf Z\parallel a_3}, $ ${\bf X\parallel a_1\parallel \Gamma-M}$, and ${\bf Y\perp X,Z}. $ Blue (orange) dots refer to Nb (Se) atoms. (d)Phonon dispersions. Red dots refer to the phonons considered.}
\label{fig:9}
\end{figure*}

In the two longitudinal modes 3 and 9, the clustered phonon displacements are similar, differing only in the phase of the Se displacements with respect to Nb: when Nb moves toward the line of Se atoms, the Se atoms move toward (away from) the Nb plane in mode 9 (3), Fig.~\ref{fig:9}, so the change in eigenfunctions upon anticrossing amounts merely to flipping the Se phase. In the intermediate mode 6 the Se is rotated from near-Z to near-Y as it moves toward Z.

We note that time molecules such as Na$_3$ are chiral.  However, a solid which is composed of many orbitals, can only be chiral if there is an imbalance between the clockwise and anticlockwise rotating molecules. Therefore, bulk NbSe$_2$, which is centrosymmetric, cannot be chiral, but chirality is allowed in  monolayer NbSe$_2$.

\subsubsection{Beyond the Born-Oppenheimer (BO) approximation}
The EPW code, see \nameref{SM3}, includes a perturbative nonadiabatic (i.e., beyond BO) correction, which we have found to be small in NbSe$_2$, although nonadiabatic effects can be much stronger.\cite{nonadi,BBO1,BBO2} Since the BO approximation relies on the large difference in the time scales involved in the electronic and ionic motions, nonadiabatic corrections would be larger when electrons move slowly as in heavy Fermion materials, flat bands, and near van Hove singularities (VHSs). Optical phonons can have strong nonadiabatic effects, as electron hoppings can be strongly enhanced for certain phonon motions, as discussed in the preceding subsection.

\subsubsection{Estimating temperature}
\begin{figure}[h] 
\includegraphics[width=\linewidth]{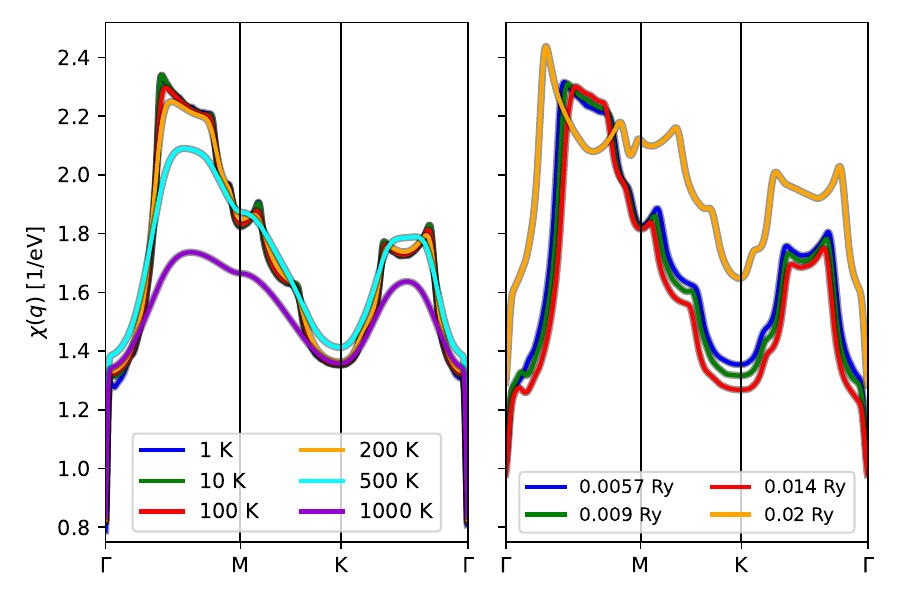}   
\vskip-0.3truecm
\caption{(color online)
Left panel: Susceptibility using the minimum $\sigma$ value at various Fermi-Dirac temperatures. Right panel: Susceptibility for fixed Fermi-Dirac temperature $T=1$~K for various $\sigma$ values.}
\label{fig:6e}
\end{figure}

The broadening parameter $\sigma$ is often introduced as a {\it fictitious temperature} to improve the convergence of the DFT self-consistency cycles towards the ground state\cite{fict}. Using the Fermi function with $\sigma=k_BT_\sigma$, however, leads to a large volume of $k$-space that must be integrated over, so that a more confined broadening function, such as a Gaussian, is a better choice. To gain a handle on relating $\sigma$ to an effective temperature, we can compare the results of $\sigma$-broadening with direct calculations of the $T$-dependent Lindhard susceptibility, Fig.~\ref{fig:6e}. Focusing on the broadening of the susceptibility peak along $\Gamma - M - K$, it is seen to be very small up to $\sigma=0.016$~Ry ($T_{\sigma}=2526$~K) where the peak is less reduced than for $T=250K$. But, the $\sigma=0.02$ susceptibility ($T_{\sigma}=3157$K) is much more intense and broader than at other $\sigma$ values, with a peak-width that is comparable to the $T=1000$K susceptibility. These results suggest that $T_{\sigma}/T$ is  $\leq 10$ and $\geq 3.5$. Since the CDW transition occurs at $\sigma=0.0057$~Ry, or $T_{\sigma}=900$K, we thus estimate that the corresponding $T_{CDW}$ lies between 225K and 90K, which is consistent with the experimental value for monolayer NbSe$_2$, $T_{CDW}=145$K.
 
\section{Conclusions}
We have developed a technique to uncover the quantum strangeness associated with anti-crossing phenomena and their important role in the softening of optical phonons.  An in-depth study of the giant Kohn anomaly and circular phonons in 1H-NbSe$_2$ transition-metal-dichalcogenide is presented as an illustrative application of our technique. We provide compelling evidence that the soft phonon in NbSe$_2$ is a longitudinal optical phonon which softens by anti-crossing several intervening phonon bands.  We also show that the CDW vector in NbSe$_2$ results from the convolution of the susceptibility and electron-phonon coupling, and that the softened phonons are circularly polarized. Such phonons may be relevant for generating time crystals, reducing decoherence in solid-state defect-spin qubits, among other potential applications.  Our calculations required an extension of the EPW code to compute off-diagonal mode-mode couplings in the phonon self-energy.  Our study will impact the modeling of charge density waves and other phenomena involving electron-phonon couplings by enabling accurate, material-specific computation of phonon self-energies in wide classes of materials. 

\section{Methods}

We performed first-principles calculations based on density functional theory (DFT) using the Quantum ESPRESSO package \cite{QE_refs1,QE_refs2}. The monolayer 1H-NbSe$_2$ structure was optimized with a vacuum spacing of 10~Å. We used a plane-wave basis set with a wavefunction cutoff of 100~Ry and a charge density cutoff of 800~Ry. The energy and force convergence thresholds were set to $10^{-6}$ and $10^{-5}$~a.u., respectively. We employed ultrasoft pseudopotentials for Nb and norm-conserving pseudopotentials for Se, following the methodology from the literature.~\cite{anharm}. The self-consistent field (SCF) calculations were performed on a $36 \times 36 \times 1$ Monkhorst-Pack $k$-point grid.

Phonon spectra and electron-phonon coupling matrix elements were calculated using density functional perturbation theory (DFPT) \cite{Baroni}. The phonon frequencies were computed on a $12 \times 12 \times 1$ $q$-point mesh with a precision of $10^{-16}$~Ry. The electron-phonon coupling (EPC) matrix was evaluated using the EPW code \cite{Giustino1} along a $Q$-path sampled with 500 points, utilizing a dense $500 \times 500 \times 1$ $k$-point mesh. To account for temperature effects, we used Fermi-Dirac smearing with a broadening parameter $\sigma$, related to an effective temperature $T_\sigma = \sigma/k_B$.

To investigate the role of phonon mode coupling, we extended the EPW code to compute the off-diagonal elements of the phonon self-energy $\Pi_{\lambda\lambda'}$. The adiabatic self-energy $\Pi'_{\lambda\lambda'}$ was calculated as defined in the cited paper.~\cite{Giustino1}. The dressed phonon frequencies $\Omega_q$ were obtained from DFPT, and the bare phonon frequencies $\omega_q$ were extracted by inverting the dynamical matrix, as discussed in the Supporting Information (see \nameref{SM3} and \nameref{SM4}).

\section*{Acknowledgements} 
The work at Northeastern University was supported by the National Science Foundation through the Expand-QISE award NSF-OMA-2329067 and benefited from the resources of Northeastern University’s Advanced Scientific Computation Center, the Discovery Cluster, the Massachusetts Technology Collaborative award MTC-22032, and the Quantum Materials and Sensing Institute. S.C. and S.E. acknowledge the support provided by the National Science Foundation through the ExpandQISE award No. 2329087. S.M. was supported by the U.S. Department of Energy (DOE), Office of Science, and Basic Energy Sciences under Award No. DE-SC0022216. The calculations performed by T.H. and C.S. in this paper were performed at TUBITAK ULAKBIM, High Performance and Grid Computing Center (TRUBA resources). The research at Howard University used the resources of Accelerate ACCESS  PHYS220127 and PHYS2100073. We thank Adrian Feiguin for stimulating discussions. T.H. is thankful to  Cagliyan Kurdak and the Department of Physics of the University of Michigan for partial support. 

\section{Associated Content}
A preprint version of this work is available as Susy Exists, Sougata Mardanya, Robert Markiewicz, Tugrul Hakioglu, Jouko Nieminen, Ville J. Härkönen, Cem Sanga, Arun Bansil, and Sugata Chowdhury,
Giant Kohn anomaly and chiral phonons in the charge density wave phase of 1H-NbSe$_2$, 2025, arXiv:2501.04821, arXiv, 
https://doi.org/10.48550/arXiv.2501.04821 (accessed 7 January, 2026).

\textbf{Supporting Information Available:} Toy model of Kohn ladders; k-resolved deconvolution of the phonon self-energy; DFPT/self-energy formalism; inverse dynamical matrix approach for extracting bare phonon frequencies; DFT/DFPT/EPC computational parameters and pseudopotentials; Kekul\'e-type $K$-point phonon displacement patterns in graphene.

\bibliography{biblio}

@book{Pauling,
  title={The Nature of the Chemical Bond and the Structure of Molecules and Crystals: An Introduction to Modern Structural Chemistry},
  author={Pauling, L.},
  isbn={9780801403330},
  lccn={60016025},
  series={George Fisher Baker Non-Resident Lecture Series},
  url={https://books.google.com/books?id=L-1K9HmKmUUC},
  year={1960},
  publisher={Cornell University Press}
}

@article{TMDC1,
	title = {{2D} transition metal dichalcogenides},
	volume = {2},
	copyright = {2017 Macmillan Publishers Limited},
	doi = {10.1038/natrevmats.2017.33},
	number = {8},
	journal = {Nat Rev Mater},
	author = {Manzeli, Sajedeh and Ovchinnikov, Dmitry and Pasquier, Diego and Yazyev, Oleg V. and Kis, Andras},
	month = jun,
	year = {2017},
}

@article{TMDC2,
	title = {Two-dimensional layered materials: {Structure}, properties, and prospects for device applications},
	volume = {29},
	doi = {10.1557/jmr.2014.6},
	number = {3},
	journal = {Journal of Materials Research},
	author = {Kaul, Anupama B.},
	month = feb,
	year = {2014},
	pages = {348--361},
}

@article{TMDC3,
	title = {Nanoscale {Transition} {Metal} {Dichalcogenides}: {Structures}, {Properties}, and {Applications}},
	volume = {39},
	doi = {10.1080/10408436.2013.863176},
	number = {5},
	journal = {Critical Reviews in Solid State and Materials Sciences},
	author = {Sorkin, V. and Pan, H. and Shi, H. and Quek, S. Y. and Zhang, Y. W.},
	month = sep,
	year = {2014},
	pages = {319--367},
}

@book{Mahan,
  author = {Mahan, G. D.},
  title = {Many-Particle Physics},
  publisher = {Springer},
  edition = {3rd},
  year = {2000},
  isbn = {978-0306463389},
  doi = {10.1007/978-1-4757-5714-9}
}

@article{Giustino2017,
  author = {Giustino, F.},
  title = {Electron-phonon interactions from first principles},
  journal = {Reviews of Modern Physics},
  volume = {89},
  number = {1},
  pages = {015003},
  year = {2017},
  doi = {10.1103/RevModPhys.89.015003}
}

@article{Marini2001,
  author = {Marini, A. and Del Sole, R.},
  title = {Possibility of a microscopic description of electron-phonon interaction with density-functional theory},
  journal = {Physical Review Letters},
  volume = {87},
  number = {27},
  pages = {276403},
  year = {2001},
  doi = {10.1103/PhysRevLett.87.276403}
}

@article{Heid2010,
  author = {Heid, R. and Bohnen, K.-P. and Zeyher, R. and Manske, D.},
  title = {Momentum dependence of the electron-phonon coupling and self-energy effects in superconducting YBa$_2$Cu$_3$O$_7$ within the local-density approximation},
  journal = {Physical Review B},
  volume = {81},
  pages = {174527},
  year = {2010},
  doi = {10.1103/PhysRevB.81.174527}
}

@article{Harkonen2020,
  author = {H\"ark\"onen, V. J. and van Leeuwen, R. and Gross, E. K. U.},
  title = {Many-body Green's function theory of electrons and nuclei beyond the {Born-Oppenheimer} approximation},
  journal = {Physical Review B},
  volume = {101},
  pages = {235153},
  year = {2020},
  doi = {10.1103/PhysRevB.101.235153}
}

@article{EPH3,
    author = {F. Weber and S. Rosenkranz and J.-P. Castellan and R. Osborn and R. Hott and R. Heid and K.-P. Bohnen and T. Egami and A.~H. Said and D. Reznik},
    title = {Extended phonon collapse and the origin of the charge-density wave in {2H-NbSe}\textsubscript{2}},
    journal = {Physical Review Letters},
    volume = {107},
    pages = {107403},
    year = {2011},
    doi = {10.1103/PhysRevLett.107.107403}
}

@article{Soumyanarayanan2013,
  author = {Soumyanarayanan, A. and Yee, M. M. and He, Y. and van Wezel, J. and Rahn, D. J. and Rossnagel, K. and Hudson, E. W. and Norman, M. R. and Hoffman, J. E.},
  title = {Quantum phase transition from triangular to stripe charge order in {NbSe$_2$}},
  journal = {Proceedings of the National Academy of Sciences},
  volume = {110},
  number = {5},
  pages = {1623--1627},
  year = {2013},
  doi = {10.1073/pnas.1211387110}
}

@article{magnetic,
title = {The low-temperature electrical and magnetic properties of TaSe2 and {NbSe$_2$}},
journal = {Journal of Solid State Chemistry},
volume = {1},
number = {2},
pages = {190-194},
year = {1970},
issn = {0022-4596},
doi = {https://doi.org/10.1016/0022-4596(70)90013-7},
author = {H.N.S. Lee and M. Garcia and H. McKinzie and A. Wold},
}

@article{Johannes2006,
  author = {Johannes, M. D. and Mazin, I. I.},
  title = {Fermi surface nesting and the origin of charge density waves in metals},
  journal = {Physical Review B},
  volume = {77},
  pages = {165135},
  year = {2008},
  doi = {10.1103/PhysRevB.77.165135}
}

@article{Borisenko2009,
  author = {Borisenko, S. V. and Kordyuk, A. A. and Zabolotnyy, V. B. and Evtushinsky, D. V. and B{\"u}chner, B. and Yaresko, A. N. and Follath, R.},
  title = {Two energy gaps and Fermi-surface ``arcs'' in {NbSe$_2$}},
  journal = {Physical Review Letters},
  volume = {102},
  pages = {166402},
  year = {2009},
  doi = {10.1103/PhysRevLett.102.166402}
}

@article{Kiss2007,
  author = {Kiss, T. and Yokoya, T. and Chainani, A. and Shin, S. and Hanaguri, T. and Nohara, M. and Takagi, H.},
  title = {Charge-order-maximized momentum-dependent superconductivity},
  journal = {Nature Physics},
  volume = {3},
  pages = {720--725},
  year = {2007},
  doi = {10.1038/nphys699}
}

@article{Calandra2009,
  author = {Calandra, M. and Mauri, F.},
  title = {Charge-density wave and superconducting dome in TiSe$_2$ from electron-phonon interaction},
  journal = {Physical Review Letters},
  volume = {101},
  pages = {016404},
  year = {2008},
  doi = {10.1103/PhysRevLett.101.016404}
}

@article{Wilson1975,
  author = {Wilson, J. A. and Di Salvo, F. J. and Mahajan, S.},
  title = {Charge-Density Waves in Metallic, Layered, Transition-Metal Dichalcogenides},
  journal = {Advances in Physics},
  volume = {24},
  number = {2},
  pages = {117--201},
  year = {1975},
  doi = {10.1080/00018737500101391}
}

@article{Moncton1977,
  author = {Moncton, D. E. and Axe, J. D. and DiSalvo, F. J.},
  title = {Neutron Scattering Study of the Charge-Density Wave Transitions in {2H-NbSe\textsubscript{2}} and {2H-TaSe\textsubscript{2}}},
  journal = {Physical Review B},
  volume = {16},
  number = {2},
  pages = {801--819},
  year = {1977},
  doi = {10.1103/PhysRevB.16.801}
}

@article{Na3,
  title = {Fractional Quantization of Molecular Pseudorotation in ${\mathrm{Na}}_{3}$},
  author = {Delacr\'etaz, Guy and Grant, Edward R. and Whetten, Robert L. and W\"oste, Ludger and Zwanziger, Josef W.},
  journal = {Phys. Rev. Lett.},
  volume = {56},
  issue = {24},
  pages = {2598--2601},
  year = {1986},
  month = {Jun},
  publisher = {American Physical Society},
  doi = {10.1103/PhysRevLett.56.2598},
}

@article{LongHi,
author = {Longuet-Higgins, Hugh Christopher  and Öpik, U.  and Pryce, Maurice Henry Lecorney  and Sack, R. A. },
title = {Studies of the {Jahn-Teller} effect .II. The dynamical problem},
journal = {Proceedings of the Royal Society of London. Series A. Mathematical and Physical Sciences},
volume = {244},
number = {1236},
pages = {1-16},
year = {1958},
doi = {10.1098/rspa.1958.0022},
}

@incollection{nonadi,
  author    = {Lundqvist, Bengt I. and Hellman, Anders and Zori{\'c}, Igor},
  title     = {Electron Transfer and Nonadiabaticity},
  booktitle = {Dynamics},
  editor    = {Hasselbrink, Eckart and Lundqvist, Bengt I.},
  series    = {Handbook of Surface Science},
  edition   = {1},
  publisher = {North-Holland},
  address   = {Amsterdam, The Netherlands},
  year      = {2008},
  volume    = {3},
  pages     = {429--524},
  doi       = {10.1016/S1573-4331(08)00010-3}
}

@book{BBO1,
author = {Baer, M. (Michael)},
address = {Hoboken, N.J},
booktitle = {Beyond {Born-Oppenheimer} : conical intersections and electronic nonadiabatic coupling terms},
isbn = {0471778915},
lccn = {2005021350},
publisher = {Wiley},
title = {Beyond {Born-Oppenheimer} : conical intersections and electronic nonadiabatic coupling terms / by Michael Baer.},
year = {2006},
}

@Article{BBO2,
author={Pisana, Simone
and Lazzeri, Michele
and Casiraghi, Cinzia
and Novoselov, Kostya S.
and Geim, A. K.
and Ferrari, Andrea C.
and Mauri, Francesco},
title={Breakdown of the adiabatic {Born--Oppenheimer} approximation in graphene},
journal={Nature Materials},
year={2007},
month={Mar},
day={01},
volume={6},
number={3},
pages={198-201},
issn={1476-4660},
doi={10.1038/nmat1846},
}

@article{CuprateSoften1,
  title = {Quantum Fluctuations of Charge Order Induce Phonon Softening in a Superconducting Cuprate},
  author = {Huang, H. Y. and Singh, A. and Mou, C. Y. and Johnston, S. and Kemper, A. F. and van den Brink, J. and Chen, P. J. and Lee, T. K. and Okamoto, J. and Chu, Y. Y. and Li, J. H. and Komiya, S. and Komarek, A. C. and Fujimori, A. and Chen, C. T. and Huang, D. J.},
  journal = {Phys. Rev. X},
  volume = {11},
  issue = {4},
  pages = {041038},
  numpages = {9},
  year = {2021},
  month = {Nov},
  publisher = {American Physical Society},
  doi = {10.1103/PhysRevX.11.041038}}

@article{CuprateSoften2,
  title = {Anomalous softening of phonon dispersion in cuprate superconductors},
  author = {Sarkar, Saheli and Grandadam, Maxence and P\'epin, Catherine},
  journal = {Phys. Rev. Res.},
  volume = {3},
  issue = {1},
  pages = {013162},
  numpages = {12},
  year = {2021},
  month = {Feb},
  publisher = {American Physical Society},
  doi = {10.1103/PhysRevResearch.3.013162},
  url = {https://link.aps.org/doi/10.1103/PhysRevResearch.3.013162}
}

@book{OD,
  title={Structural Phase Transitions},
  author={Bruce, A.D. and Cowley, R.A.},
  isbn={9780850662061},
  lccn={81150512},
  series={Monographs on Physics: Taylor and Francis},
  url={https://books.google.com/books?id=JPFAAQAAIAAJ},
  year={1981},
  publisher={Taylor \& Francis}
}

@article{CuprateSoften3,
doi = {10.1088/2515-7639/ad6c7f},
url = {https://dx.doi.org/10.1088/2515-7639/ad6c7f},
year = {2024},
month = {aug},
publisher = {IOP Publishing},
volume = {7},
number = {4},
pages = {045002},
author = {Cunyuan Jiang and Giovanni Alberto Ummarino and Matteo Baggioli and Efthymios Liarokapis and Alessio Zaccone},
title = {Correlation between optical phonon softening and superconducting $T_\mathrm{c}$ in {YBa2Cu3Ox} within d-wave Eliashberg theory},
journal = {Journal of Physics: Materials}
}

@article{fict,
  title = {Fermi energy determination for advanced smearing techniques},
  author = {dos Santos, Flaviano Jos\'e and Marzari, Nicola},
  journal = {Phys. Rev. B},
  volume = {107},
  issue = {19},
  pages = {195122},
  numpages = {10},
  year = {2023},
  month = {May},
  publisher = {American Physical Society},
  doi = {10.1103/PhysRevB.107.195122},
}

@article{Plummer,
   title={Classification of charge density waves based on their nature},
   volume={112},
   ISSN={1091-6490},
   url={http://dx.doi.org/10.1073/pnas.1424791112},
   DOI={10.1073/pnas.1424791112},
   number={8},
   journal={Proceedings of the National Academy of Sciences},
   publisher={Proceedings of the National Academy of Sciences},
   author={Zhu, Xuetao and Cao, Yanwei and Zhang, Jiandi and Plummer, E. W. and Guo, Jiandong},
   year={2015},
   month=feb, pages={2367–2371} }

@article{time1,
  title = {Quantum Time Crystals},
  author = {Wilczek, Frank},
  journal = {Phys. Rev. Lett.},
  volume = {109},
  issue = {16},
  pages = {160401},
  numpages = {5},
  year = {2012},
  month = {Oct},
  publisher = {American Physical Society},
  doi = {10.1103/PhysRevLett.109.160401},
}

@article{time2,
   title={Time crystals: a review},
   volume={81},
   ISSN={1361-6633},
   DOI={10.1088/1361-6633/aa8b38},
   number={1},
   journal={Reports on Progress in Physics},
   publisher={IOP Publishing},
   author={Sacha, Krzysztof and Zakrzewski, Jakub},
   year={2017},
   month=nov, pages={016401} }

@article{Berciu,
  title = {Peierls versus Holstein models for describing electron-phonon coupling in perovskites},
  author = {Yam, Yau-Chuen and Moeller, Mirko M. and Sawatzky, George A. and Berciu, Mona},
  journal = {Phys. Rev. B},
  volume = {102},
  issue = {23},
  pages = {235145},
  numpages = {14},
  year = {2020},
  month = {Dec},
  publisher = {American Physical Society},
  doi = {10.1103/PhysRevB.102.235145},
}

@article{MIT,
  title = {Metal-Insulator Transition},
  author = {Mott, N. F.},
  journal = {Rev. Mod. Phys.},
  volume = {40},
  issue = {4},
  pages = {677--683},
  numpages = {0},
  year = {1968},
  month = {Oct},
  publisher = {American Physical Society},
  doi = {10.1103/RevModPhys.40.677},
}

@book{Mott,
author = {Alexandrov, A S and Mott, N F},
title = {Polarons and Bipolarons},
publisher = {World Scientific},
year = {1996},
doi = {10.1142/2784},
address = {},
edition   = {},
}

@article{Kohn,
  title = {Image of the Fermi Surface in the Vibration Spectrum of a Metal},
  author = {Kohn, W.},
  journal = {Phys. Rev. Lett.},
  volume = {2},
  issue = {9},
  pages = {393--394},
  numpages = {0},
  year = {1959},
  month = {May},
  publisher = {American Physical Society},
  doi = {10.1103/PhysRevLett.2.393},
}

@article{Gutzcharge,
doi = {10.1088/1367-2630/17/2/023074},
year = {2015},
month = {feb},
publisher = {IOP Publishing},
volume = {17},
number = {2},
pages = {023074},
author = {R S Markiewicz and G Seibold and J Lorenzana and A Bansil},
title = {Gutzwiller charge phase diagram of cuprates, including electron–phonon coupling effects},
journal = {New Journal of Physics},
}

@article{Ding,
author = {Jiemin Li  and Abhishek Nag  and Jonathan Pelliciari  and Hannah Robarts  and Andrew Walters  and Mirian Garcia-Fernandez  and Hiroshi Eisaki  and Dongjoon Song  and Hong Ding  and Steven Johnston  and Riccardo Comin  and Ke-Jin Zhou },
title = {Multiorbital charge-density wave excitations and concomitant phonon anomalies in {Bi$_{2}$Sr$_{2}$LaCuO$_{6+\delta}$}},
journal = {Proceedings of the National Academy of Sciences},
volume = {117},
number = {28},
pages = {16219-16225},
year = {2020},
doi = {10.1073/pnas.2001755117},
}

@article{Forgan,
author={Forgan, E. M.
and Blackburn, E.
and Holmes, A. T.
and Briffa, A. K. R.
and Chang, J.
and Bouchenoire, L.
and Brown, S. D.
and Liang, Ruixing
and Bonn, D.
and Hardy, W. N.
and Christensen, N. B.
and Zimmermann, M. V.
and H{\"u}cker, M.
and Hayden, S. M.},
title={The microscopic structure of charge density waves in underdoped {YBa$_2$Cu$_3$O${}_{6.54}$} revealed by X-ray diffraction},
journal={Nature Communications},
year={2015},
month={Dec},
day={09},
volume={6},
number={1},
pages={10064},
doi={10.1038/ncomms10064},
}

@article{Rez,
  title = {Optical phonons and the soft mode in {2$H$-NbSe${}_{2}$}},
  author = {Weber, F. and Hott, R. and Heid, R. and Bohnen, K.-P. and Rosenkranz, S. and Castellan, J.-P. and Osborn, R. and Said, A. H. and Leu, B. M. and Reznik, D.},
  journal = {Phys. Rev. B},
  volume = {87},
  issue = {24},
  pages = {245111},
  numpages = {8},
  year = {2013},
  month = {Jun},
  publisher = {American Physical Society},
  doi = {10.1103/PhysRevB.87.245111},
  url = {https://link.aps.org/doi/10.1103/PhysRevB.87.245111}
}

@article{NbSe2,
author = {Chiu, Wei-Chi and Mardanya, Sougata and Markiewicz, Robert and Nieminen, Jouko and Singh, Bahadur and Hakioglu, Tugrul and Agarwal, Amit and Chang, Tay-Rong and Lin, Hsin and Bansil, Arun},
title = {Strain-Induced Charge Density Waves with Emergent Topological States in Monolayer NbSe2},
journal = {ACS Nano},
volume = {19},
number = {19},
pages = {18108-18116},
year = {2025},
doi = {10.1021/acsnano.4c13478},
note ={PMID: 40327834},
URL = {https://doi.org/10.1021/acsnano.4c13478},
eprint = {https://doi.org/10.1021/acsnano.4c13478}
}

@article{Flick,
   title={Charge order in {NbSe$_2$}},
   volume={94},
   ISSN={2469-9969},
   url={http://dx.doi.org/10.1103/PhysRevB.94.235135},
   DOI={10.1103/physrevb.94.235135},
   number={23},
   journal={Physical Review B},
   publisher={American Physical Society (APS)},
   author={Flicker, Felix and van Wezel, Jasper},
   year={2016},
   month=dec }

@article{anharm,
   title={Weak Dimensionality Dependence and Dominant Role of Ionic Fluctuations in the Charge-Density-Wave Transition of 
{NbSe{$_2$}}},
   volume={125},
   ISSN={1079-7114},
   url={http://dx.doi.org/10.1103/PhysRevLett.125.106101},
   DOI={10.1103/physrevlett.125.106101},
   number={10},
   journal={Physical Review Letters},
   publisher={American Physical Society (APS)},
   author={Bianco, Raffaello and Monacelli, Lorenzo and Calandra, Matteo and Mauri, Francesco and Errea, Ion},
   year={2020},
   month=sep }

@article{Giustino1,
   title={Electron-phonon interactions from first principles},
   volume={89},
   DOI={10.1103/revmodphys.89.015003},
   journal={Reviews of Modern Physics},
   author={Giustino, Feliciano},
   year={2017},
}

@book{Grimvall,
author={Grimvall, Göran},
title={The electron-phonon interaction in metals},
series={Series of monographs on selected topics in solid state physics; 16},
year={1981},
publisher={North-Holland Pub. Co.},
isbn={044486105X},
}

@article{Baroni,
  title = {Phonons and related crystal properties from density-functional perturbation theory},
  author = {Baroni, Stefano and de Gironcoli, Stefano and Dal Corso, Andrea and Giannozzi, Paolo},
  journal = {Rev. Mod. Phys.},
  volume = {73},
  issue = {2},
  pages = {515--562},
  numpages = {0},
  year = {2001},
  month = {Jul},
  publisher = {American Physical Society},
  doi = {10.1103/RevModPhys.73.515},
  url = {https://link.aps.org/doi/10.1103/RevModPhys.73.515}
}

@inbook{Hedin,
title = "Effects of Electron-Electron and Electron-Phonon Interactions on the One-Electron States of Solids",
author = "Lars Hedin and Lundqvist, {Stig O.}",
note = "doi:10.1016/S0081-1947(08)60615-3",
year = "1969",
doi = "10.1016/S0081-1947(08)60615-3",
isbn = "978-0-12-607723-0",
volume = "23",
publisher = "Academic Press",
pages = "1--181",
editor = "Frederick Seitz and David Turnbull and Henry Ehrenreich",
booktitle = "Solid State Physics",
address = "United States",
}

@article{Wannier,
doi = {10.1088/1361-648X/ab51ff},
url = {https://dx.doi.org/10.1088/1361-648X/ab51ff},
year = {2020},
month = {jan},
publisher = {IOP Publishing},
volume = {32},
number = {16},
pages = {165902},
author = {Giovanni Pizzi and Valerio Vitale and Ryotaro Arita and Stefan Blügel and Frank Freimuth and Guillaume Géranton and Marco Gibertini and Dominik Gresch and Charles Johnson and Takashi Koretsune and Julen Ibañez-Azpiroz and Hyungjun Lee and Jae-Mo Lihm and Daniel Marchand and Antimo Marrazzo and Yuriy Mokrousov and Jamal I Mustafa and Yoshiro Nohara and Yusuke Nomura and Lorenzo Paulatto and Samuel Poncé and Thomas Ponweiser and Junfeng Qiao and Florian Thöle and Stepan S Tsirkin and Małgorzata Wierzbowska and Nicola Marzari and David Vanderbilt and Ivo Souza and Arash A Mostofi and Jonathan R Yates},
title = {Wannier90 as a community code: new features and applications},
journal = {Journal of Physics: Condensed Matter},
}

@article{MatBar,
  title = {Theory of the Anomalous Skin Effect in Normal and Superconducting Metals},
  author = {Mattis, D. C. and Bardeen, J.},
  journal = {Phys. Rev.},
  volume = {111},
  issue = {2},
  pages = {412--417},
  numpages = {0},
  year = {1958},
  month = {Jul},
  publisher = {American Physical Society},
  doi = {10.1103/PhysRev.111.412},
  url = {https://link.aps.org/doi/10.1103/PhysRev.111.412}
}

@article{Joh,
  title = {Second dome of superconductivity in ${\mathrm{YBa}}_{2}{\mathrm{Cu}}_{3}{\mathrm{O}}_{7}$ at high pressure},
  author = {Nokelainen, Johannes and Matzelle, Matthew E. and Lane, Christopher and Atlam, Nabil and Zhang, Ruiqi and Markiewicz, Robert S. and Barbiellini, Bernardo and Sun, Jianwei and Bansil, Arun},
  journal = {Phys. Rev. B},
  volume = {110},
  issue = {2},
  pages = {L020502},
  numpages = {6},
  year = {2024},
  month = {Jul},
  publisher = {American Physical Society},
  doi = {10.1103/PhysRevB.110.L020502},
  url = {https://link.aps.org/doi/10.1103/PhysRevB.110.L020502}
}

@article{Zak,
  title = {Connectivity of energy bands in crystals},
  author = {Michel, L. and Zak, J.},
  journal = {Phys. Rev. B},
  volume = {59},
  issue = {9},
  pages = {5998--6001},
  numpages = {0},
  year = {1999},
  month = {Mar},
  publisher = {American Physical Society},
  doi = {10.1103/PhysRevB.59.5998},
  url = {https://link.aps.org/doi/10.1103/PhysRevB.59.5998}
}

@article{graphene,
  title = {Nonadiabatic {Kohn} Anomaly in a Doped Graphene Monolayer},
  author = {Lazzeri, Michele and Mauri, Francesco},
  journal = {Phys. Rev. Lett.},
  volume = {97},
  issue = {26},
  pages = {266407},
  numpages = {4},
  year = {2006},
  month = {Dec},
  publisher = {American Physical Society},
  doi = {10.1103/PhysRevLett.97.266407},
  url = {https://link.aps.org/doi/10.1103/PhysRevLett.97.266407}
}

@article{
SLR1,
author = {Alessandro Lunghi  and Stefano Sanvito },
title = {How do phonons relax molecular spins?},
journal = {Science Advances},
volume = {5},
number = {9},
year = {2019},
doi = {10.1126/sciadv.aax7163},
}

@article{Mingda,
  title = {Topological Singularity Induced Chiral {Kohn} Anomaly in a {Weyl} Semimetal},
  author = {Nguyen, Thanh and Han, Fei and Andrejevic, Nina and Pablo-Pedro, Ricardo and Apte, Anuj and Tsurimaki, Yoichiro and Ding, Zhiwei and Zhang, Kunyan and Alatas, Ahmet and Alp, Ercan E. and Chi, Songxue and Fernandez-Baca, Jaime and Matsuda, Masaaki and Tennant, David Alan and Zhao, Yang and Xu, Zhijun and Lynn, Jeffrey W. and Huang, Shengxi and Li, Mingda},
  journal = {Phys. Rev. Lett.},
  volume = {124},
  issue = {23},
  pages = {236401},
  numpages = {7},
  year = {2020},
  month = {Jun},
  publisher = {American Physical Society},
  doi = {10.1103/PhysRevLett.124.236401},
  url = {https://link.aps.org/doi/10.1103/PhysRevLett.124.236401}
}

@article{QE_refs1,
doi = {10.1088/0953-8984/21/39/395502},
url = {https://dx.doi.org/10.1088/0953-8984/21/39/395502},
year = {2009},
month = {sep},
publisher = {},
volume = {21},
number = {39},
pages = {395502},
author = {Paolo Giannozzi and Stefano Baroni and Nicola Bonini and Matteo Calandra and Roberto Car and Carlo Cavazzoni and Davide Ceresoli and Guido L Chiarotti and Matteo Cococcioni and Ismaila Dabo and Andrea Dal Corso and Stefano de Gironcoli and Stefano Fabris and Guido Fratesi and Ralph Gebauer and Uwe Gerstmann and Christos Gougoussis and Anton Kokalj and Michele Lazzeri and Layla Martin-Samos and Nicola Marzari and Francesco Mauri and Riccardo Mazzarello and Stefano Paolini and Alfredo Pasquarello and Lorenzo Paulatto and Carlo Sbraccia and Sandro Scandolo and Gabriele Sclauzero and Ari P Seitsonen and Alexander Smogunov and Paolo Umari and Renata M Wentzcovitch},
title = {QUANTUM ESPRESSO: a modular and open-source software project for quantum
simulations of materials},
journal = {Journal of Physics: Condensed Matter},
}

@article{QE_refs2,
   title={Advanced capabilities for materials modelling with Quantum ESPRESSO},
   volume={29},
   ISSN={1361-648X},
   url={http://dx.doi.org/10.1088/1361-648X/aa8f79},
   DOI={10.1088/1361-648x/aa8f79},
   number={46},
   journal={Journal of Physics: Condensed Matter},
   publisher={IOP Publishing},
   author={Giannozzi, P and Andreussi, O and Brumme, T and Bunau, O and Buongiorno Nardelli, M and Calandra, M and Car, R and Cavazzoni, C and Ceresoli, D and Cococcioni, M and Colonna, N and Carnimeo, I and Dal Corso, A and de Gironcoli, S and Delugas, P and DiStasio, R A and Ferretti, A and Floris, A and Fratesi, G and Fugallo, G and Gebauer, R and Gerstmann, U and Giustino, F and Gorni, T and Jia, J and Kawamura, M and Ko, H-Y and Kokalj, A and Küçükbenli, E and Lazzeri, M and Marsili, M and Marzari, N and Mauri, F and Nguyen, N L and Nguyen, H-V and Otero-de-la-Roza, A and Paulatto, L and Poncé, S and Rocca, D and Sabatini, R and Santra, B and Schlipf, M and Seitsonen, A P and Smogunov, A and Timrov, I and Thonhauser, T and Umari, P and Vast, N and Wu, X and Baroni, S},
   year={2017},
   month=oct, pages={465901} }

@misc{QuantEspRenorm,
  author    = {{Quantum Espresso Foundation}},
  title     = {Phonon User's Guide},
  url       = {https://www.quantum-espresso.org/Doc/user_guide_PDF/ph_user_guide.pdf},
  urldate   = {2026-01-20},
  note      = {(accessed January 20, 2026).}
}

@misc{EPWmodified,
  author    = {Exists, Susy},
  title     = {q-e. GitLab repository (modified file)},
  url       = {https://gitlab.com/susyexists/q-e},
  urldate   = {2025-12-15},
  note      = {(accessed December 15, 2025).}
}

@misc{typy,
  author    = {Exists, Susy},
  title     = {typy. GitHub repository},
  url       = {https://github.com/susyexists/typy},
  urldate   = {2025-12-15},
  note      = {(accessed December 15, 2025).}
}

@misc{phonon_visualizer,
  author    = {Kumbhar, Snehal and Pizzi, Giovanni and Sohier, Thibault and Miranda, Henrique},
  title     = {Interactive phonon visualizer},
  url       = {https://interactivephonon.materialscloud.io/},
  urldate   = {2025-12-15},
  note      = {(accessed December 15, 2025).}
}

@article{FengZheng,
	title = {Electron-phonon coupling and the coexistence of superconductivity and charge-density wave in monolayer {NbSe$_2$}},
	volume = {99},
	issn = {2469-9950, 2469-9969},
	url = {https://link.aps.org/doi/10.1103/PhysRevB.99.161119},
	doi = {10.1103/PhysRevB.99.161119},
	number = {16},
	urldate = {2024-07-10},
	journal = {Phys. Rev. B},
	author = {Zheng, Feipeng and Feng, Ji},
	month = apr,
	year = {2019},
	pages = {161119},
}

@article{ScCdw2,
author = {Lian, Chao-Sheng and Si, Chen and Duan, Wenhui},
title = {Unveiling Charge-Density Wave, Superconductivity, and Their Competitive Nature in Two-Dimensional {NbSe$_2$}},
journal = {Nano Letters},
volume = {18},
number = {5},
pages = {2924-2929},
year = {2018},
doi = {10.1021/acs.nanolett.8b00237},
}

@article{ScCdw3,
    author = {M.~M. Ugeda and A.~J. Bradley and S.-F. Shi and F.~H. da Jornada and Y. Zhang and D.~Y. Qiu and W. Ruan and S.-K. Mo and Z. Hussain and Z.-X. Shen and F. Wang and S.~G. Louie and M.~F. Crommie},
    title = {Characterization of collective ground states in single-layer {NbSe}\textsubscript{2}},
    journal = {Nature Physics},
    volume = {12},
    pages = {92--97},
    year = {2016},
    doi = {10.1038/nphys3527}
}

@article{ScCdw4,
    author = {X. Xi and L. Zhao and Z. Wang and H. Berger and L. Forr\'o and J. Shan and K.~F. Mak},
    title = {Strongly enhanced charge-density-wave order in monolayer {NbSe}\textsubscript{2}},
    journal = {Nature Nanotechnology},
    volume = {10},
    pages = {765--769},
    year = {2015},
    doi = {10.1038/nnano.2015.143}
}

@Article{ScCdwExp,
author={Cho, Kyuil
and Ko{\'{n}}czykowski, M.
and Teknowijoyo, S.
and Tanatar, M. A.
and Guss, J.
and Gartin, P. B.
and Wilde, J. M.
and Kreyssig, A.
and McQueeney, R. J.
and Goldman, A. I.
and Mishra, V.
and Hirschfeld, P. J.
and Prozorov, R.},
title={Using controlled disorder to probe the interplay between charge order and superconductivity in {NbSe}\textsubscript{2}},
journal={Nature Communications},
year={2018},
month={Jul},
day={18},
volume={9},
number={1},
pages={2796},
issn={2041-1723},
doi={10.1038/s41467-018-05153-0},
}

@ARTICLE{EPW,
       author = {{Lee}, Hyungjun and {Ponc{\'e}}, Samuel and {Bushick}, Kyle and {Hajinazar}, Samad and {Lafuente-Bartolome}, Jon and {Leveillee}, Joshua and {Lian}, Chao and {Lihm}, Jae-Mo and {Macheda}, Francesco and {Mori}, Hitoshi and {Paudyal}, Hari and {Sio}, Weng Hong and {Tiwari}, Sabyasachi and {Zacharias}, Marios and {Zhang}, Xiao and {Bonini}, Nicola and {Kioupakis}, Emmanouil and {Margine}, Elena R. and {Giustino}, Feliciano},
        title = "{Electron-phonon physics from first principles using the EPW code}",
      journal = {npj Computational Mathematics},
         year = 2023,
        month = dec,
       volume = {9},
       number = {1},
          eid = {156},
        pages = {156},
          doi = {10.1038/s41524-023-01107-3},
archivePrefix = {arXiv},
       eprint = {2302.08085},
 primaryClass = {cond-mat.mtrl-sci},
       adsurl = {https://ui.adsabs.harvard.edu/abs/2023npjCM...9..156L}
}

@article{strain,
    author = {R. Rold\'an and A. Castellanos-Gomez and E. Cappelluti and F. Guinea},
    title = {Strain engineering in semiconducting two-dimensional crystals},
    journal = {Journal of Physics: Condensed Matter},
    volume = {27},
    number = {31},
    pages = {313201},
    year = {2015},
    doi = {10.1088/0953-8984/27/31/313201}
}

@article{hetero1,
    author = {A.~K. Geim and I.~V. Grigorieva},
    title = {Van der {Waals} heterostructures},
    journal = {Nature},
    volume = {499},
    pages = {419--425},
    year = {2013},
    doi = {10.1038/nature12385}
}

@article{hetero2,
  title = {Heterostructures of transition metal dichalcogenides},
  author = {Amin, B. and Singh, N. and Schwingenschl\"ogl, U.},
  journal = {Phys. Rev. B},
  volume = {92},
  issue = {7},
  pages = {075439},
  numpages = {6},
  year = {2015},
  month = {Aug},
  publisher = {American Physical Society},
  doi = {10.1103/PhysRevB.92.075439},
}

@article{Vaskivskyi2015Fast,
    author = {I. Vaskivskyi and J.~C. Buttler and D. Chesnel and V.~Y. Voroshnin and S. Brazovskii and P. Dzemidovich and V.~M. Svetin and A.~V. Didenko and A.~P. Zaykovskiy and M. Mihailovic},
    title = {Fast electronic resistance switching involving hidden charge density wave states},
    journal = {Nature Communications},
    volume = {7},
    pages = {11442},
    year = {2016},
    doi = {10.1038/ncomms11442}
}

@article{Akinwande2017Two,
    author = {D. Akinwande and N. Petrone and J. Hone},
    title = {Two-dimensional flexible nanoelectronics},
    journal = {Nature Communications},
    volume = {8},
    pages = {15295},
    year = {2017},
    doi = {10.1038/ncomms15295}
}

@article{Clarke2008Superconducting,
    author = {J. Clarke and F.~K. Wilhelm},
    title = {Superconducting quantum bits},
    journal = {Nature},
    volume = {453},
    pages = {1031--1042},
    year = {2008},
    doi = {10.1038/nature07128}
}

@article{FermiGap,
	title = {Two {Energy} {Gaps} and {Fermi}-{Surface} “{Arcs}” in {NbSe} 2},
	volume = {102},
	copyright = {http://link.aps.org/licenses/aps-default-license},
	issn = {0031-9007, 1079-7114},
	url = {https://link.aps.org/doi/10.1103/PhysRevLett.102.166402},
	doi = {10.1103/PhysRevLett.102.166402},
	number = {16},
	urldate = {2024-07-10},
	journal = {Phys. Rev. Lett.},
	author = {Borisenko, S. V. and Kordyuk, A. A. and Zabolotnyy, V. B. and Inosov, D. S. and Evtushinsky, D. and Büchner, B. and Yaresko, A. N. and Varykhalov, A. and Follath, R. and Eberhardt, W. and Patthey, L. and Berger, H.},
	month = apr,
	year = {2009},
	pages = {166402},
}

@article{Strangeness,
	title = {Phonon-induced giant linear-in-{T} resistivity in magic angle twisted bilayer graphene: {Ordinary} strangeness and exotic superconductivity},
	volume = {99},
	issn = {2469-9950, 2469-9969},
	shorttitle = {Phonon-induced giant linear-in-{T} resistivity in magic angle twisted bilayer graphene},
	url = {http://arxiv.org/abs/1811.04920},
	doi = {10.1103/PhysRevB.99.165112},
	number = {16},
	urldate = {2024-07-10},
	journal = {Phys. Rev. B},
	author = {Wu, Fengcheng and Hwang, Euyheon and Sarma, Sankar Das},
	month = apr,
	year = {2019},
	note = {arXiv:1811.04920 [cond-mat]},
	pages = {165112},
}

@book{gruner1994density,
  title={Density Waves In Solids},
  author={Gr{\"u}ner, G.},
  isbn={9780201626544},
  lccn={93032362},
  series={Frontiers in physics},
  url={https://books.google.com.tr/books?id=nmAsAAAAYAAJ},
  year={1994},
  publisher={Basic Books}
}

@Inbook{DiSalvo1977,
author="Di Salvo, F. J.",
editor="Riste, Tormod",
title="Charge Density Waves in Layered Compounds",
bookTitle="Electron-Phonon Interactions and Phase Transitions",
year="1977",
publisher="Springer US",
pages="107--136",
isbn="978-1-4615-8921-1",
doi="10.1007/978-1-4615-8921-1_5",
}

@article{polaron1,
  title = {Bipolaronic High-Temperature Superconductivity},
  author = {Zhang, C. and Sous, J. and Reichman, D. R. and Berciu, M. and Millis, A. J. and Prokof'ev, N. V. and Svistunov, B. V.},
  journal = {Phys. Rev. X},
  volume = {13},
  issue = {1},
  pages = {011010},
  numpages = {19},
  year = {2023},
  month = {Jan},
  publisher = {American Physical Society},
  doi = {10.1103/PhysRevX.13.011010},
  url = {https://link.aps.org/doi/10.1103/PhysRevX.13.011010}
}

@article{polaron2,
  title = {Competition of pairing and Peierls--charge-density-wave correlations in a two-dimensional electron-phonon model},
  author = {Scalettar, R. T. and Bickers, N. E. and Scalapino, D. J.},
  journal = {Phys. Rev. B},
  volume = {40},
  issue = {1},
  pages = {197--200},
  numpages = {0},
  year = {1989},
  month = {Jul},
  publisher = {American Physical Society},
  doi = {10.1103/PhysRevB.40.197},
  url = {https://link.aps.org/doi/10.1103/PhysRevB.40.197}
}

@article{topo1,
  title = {Emergence of topological states in a hybrid superconductor of monolayer {Pb} grown on $\mathrm{NbS}{\mathrm{e}}_{2}$},
  author = {Huang, Ze and Zhang, Zhenyu and Cui, Ping},
  journal = {Phys. Rev. B},
  volume = {109},
  issue = {22},
  pages = {224516},
  numpages = {12},
  year = {2024},
  month = {Jun},
  publisher = {American Physical Society},
  doi = {10.1103/PhysRevB.109.224516},
  url = {https://link.aps.org/doi/10.1103/PhysRevB.109.224516}
}

@Article{new_nbse2,
author={Das, S.
and Paudyal, H.
and Margine, E. R.
and Agterberg, D. F.
and Mazin, I. I.},
title={Electron-phonon coupling and spin fluctuations in the Ising superconductor {NbSe$_2$}},
journal={npj Computational Materials},
year={2023},
month={Apr},
day={26},
volume={9},
number={1},
pages={66},
issn={2057-3960},
doi={10.1038/s41524-023-01017-4},
url={https://doi.org/10.1038/s41524-023-01017-4}
}

@article{EPH4,
  title = {Scanning-tunneling-microscopy study of distortion and instability of inclined flux-line-lattice structures in the anisotropic superconductor {2H-${\mathrm{NbSe}}_{2}$}},
  author = {Hess, H. F. and Murray, C. A. and Waszczak, J. V.},
  journal = {Phys. Rev. Lett.},
  volume = {69},
  issue = {14},
  pages = {2138--2141},
  numpages = {0},
  year = {1992},
  month = {Oct},
  publisher = {American Physical Society},
  doi = {10.1103/PhysRevLett.69.2138},
  url = {https://link.aps.org/doi/10.1103/PhysRevLett.69.2138}
}

@article{EPH5,
doi = {10.1088/0256-307X/38/10/107101},
url = {https://dx.doi.org/10.1088/0256-307X/38/10/107101},
year = {2021},
month = {nov},
publisher = {Chinese Physical Society and IOP Publishing Ltd},
volume = {38},
number = {10},
pages = {107101},
author = {Xuedong Xie and Dongjing Lin and Li Zhu and Qiyuan Li and Junyu Zong and Wang Chen and Qinghao Meng and Qichao Tian and Shao-Chun Li and Xiaoxiang Xi and Can Wang and Yi Zhang},
title = {Charge Density Wave and Electron-Phonon Interaction in Epitaxial Monolayer {NbSe$_2$} Films},
journal = {Chinese Physics Letters}
}

@article{2new_nbse2,
  title = {Fermi surface of {$2H\ensuremath{-}{\mathrm{NbSe}}_{2}$} and its implications on the charge-density-wave mechanism},
  author = {Rossnagel, K. and Seifarth, O. and Kipp, L. and Skibowski, M. and Vo\ss{}, D. and Kr\"uger, P. and Mazur, A. and Pollmann, J.},
  journal = {Phys. Rev. B},
  volume = {64},
  issue = {23},
  pages = {235119},
  numpages = {6},
  year = {2001},
  month = {Nov},
  publisher = {American Physical Society},
  doi = {10.1103/PhysRevB.64.235119},
  url = {https://link.aps.org/doi/10.1103/PhysRevB.64.235119}
}

@book{Bersuker, place={Cambridge}, title={The Jahn-Teller Effect}, publisher={Cambridge University Press}, author={Bersuker, Isaac}, year={2006}}

@article{magnet2,
author = {Zhou, Yungang and Wang, Zhiguo and Yang, Ping and Zu, Xiaotao and Yang, Li and Sun, Xin and Gao, Fei},
title = {Tensile Strain Switched Ferromagnetism in Layered {NbS}\textsubscript{2} and {NbSe$_2$}},
journal = {ACS Nano},
volume = {6},
number = {11},
pages = {9727-9736},
year = {2012},
doi = {10.1021/nn303198w},
    note ={PMID: 23057936},

URL = { 
    
        https://doi.org/10.1021/nn303198w
},
eprint = { 
        https://doi.org/10.1021/nn303198w}}

@book{atom-transistor, place={Cambridge}, title={Quantum Transport: Atom to Transistor}, publisher={Cambridge University Press}, author={Datta, Supriyo}, year={2005}}

@article{wire,
  title = {Size-dependent resistivity of metallic wires in the mesoscopic range},
  author = {Steinh\"ogl, Werner and Schindler, G\"unther and Steinlesberger, Gernot and Engelhardt, Manfred},
  journal = {Phys. Rev. B},
  volume = {66},
  issue = {7},
  pages = {075414},
  numpages = {4},
  year = {2002},
  month = {Aug},
  publisher = {American Physical Society},
  doi = {10.1103/PhysRevB.66.075414},
  url = {https://link.aps.org/doi/10.1103/PhysRevB.66.075414}
}

@article{sensing,
  title = {Quantum sensing},
  author = {Degen, C. L. and Reinhard, F. and Cappellaro, P.},
  journal = {Rev. Mod. Phys.},
  volume = {89},
  issue = {3},
  pages = {035002},
  numpages = {39},
  year = {2017},
  month = {Jul},
  publisher = {American Physical Society},
  doi = {10.1103/RevModPhys.89.035002},
  url = {https://link.aps.org/doi/10.1103/RevModPhys.89.035002}
}

@book{nielsen, place={Cambridge}, title={Quantum Computation and Quantum Information: 10th Anniversary Edition}, publisher={Cambridge University Press}, author={Nielsen, Michael A. and Chuang, Isaac L.}, year={2010}}

@article{hse,
  title = {Heyd-Scuseria-Ernzerhof hybrid functional for calculating the lattice dynamics of semiconductors},
  author = {Hummer, Kerstin and Harl, Judith and Kresse, Georg},
  journal = {Phys. Rev. B},
  volume = {80},
  issue = {11},
  pages = {115205},
  numpages = {12},
  year = {2009},
  month = {Sep},
  publisher = {American Physical Society},
  doi = {10.1103/PhysRevB.80.115205},
  url = {https://link.aps.org/doi/10.1103/PhysRevB.80.115205}
}

@article{scan,
  title = {Strongly Constrained and Appropriately Normed Semilocal Density Functional},
  author = {Sun, Jianwei and Ruzsinszky, Adrienn and Perdew, John P.},
  journal = {Phys. Rev. Lett.},
  volume = {115},
  issue = {3},
  pages = {036402},
  numpages = {6},
  year = {2015},
  month = {Jul},
  publisher = {American Physical Society},
  doi = {10.1103/PhysRevLett.115.036402},
  url = {https://link.aps.org/doi/10.1103/PhysRevLett.115.036402}
}

@article{hasan-kane,
  title = {Colloquium: Topological insulators},
  author = {Hasan, M. Z. and Kane, C. L.},
  journal = {Rev. Mod. Phys.},
  volume = {82},
  issue = {4},
  pages = {3045--3067},
  numpages = {0},
  year = {2010},
  month = {Nov},
  publisher = {American Physical Society},
  doi = {10.1103/RevModPhys.82.3045},
  url = {https://link.aps.org/doi/10.1103/RevModPhys.82.3045}
}

@article{weyl-dirac,
  title = {Weyl and Dirac semimetals in three-dimensional solids},
  author = {Armitage, N. P. and Mele, E. J. and Vishwanath, Ashvin},
  journal = {Rev. Mod. Phys.},
  volume = {90},
  issue = {1},
  pages = {015001},
  numpages = {57},
  year = {2018},
  month = {Jan},
  publisher = {American Physical Society},
  doi = {10.1103/RevModPhys.90.015001},
  url = {https://link.aps.org/doi/10.1103/RevModPhys.90.015001}
}

@article{degeneracy,
  title = {Accidental Degeneracy in the Energy Bands of Crystals},
  author = {Herring, Conyers},
  journal = {Phys. Rev.},
  volume = {52},
  issue = {4},
  pages = {365--373},
  numpages = {0},
  year = {1937},
  month = {Aug},
  publisher = {American Physical Society},
  doi = {10.1103/PhysRev.52.365},
  url = {https://link.aps.org/doi/10.1103/PhysRev.52.365}
}

@article{crossing,
 ISSN = {09501207},
 URL = {http://www.jstor.org/stable/96038},
 author = {Clarence Zener},
 journal = {Proceedings of the Royal Society of London. Series A, Containing Papers of a Mathematical and Physical Character},
 number = {833},
 pages = {696--702},
 publisher = {The Royal Society},
 title = {Non-Adiabatic Crossing of Energy Levels},
 urldate = {2025-09-11},
 volume = {137},
 year = {1932}
}

@article{SSCHA,
doi = {10.1088/1361-648X/ac066b},
url = {https://dx.doi.org/10.1088/1361-648X/ac066b},
year = {2021},
month = {jul},
publisher = {IOP Publishing},
volume = {33},
number = {36},
pages = {363001},
author = {Monacelli, Lorenzo and Bianco, Raffaello and Cherubini, Marco and Calandra, Matteo and Errea, Ion and Mauri, Francesco},
title = {The stochastic self-consistent harmonic approximation: calculating vibrational properties of materials with full quantum and anharmonic effects},
journal = {Journal of Physics: Condensed Matter}
}

@article{PhysRevB.39.13120,
  title = {Density-functional approach to nonlinear-response coefficients of solids},
  author = {Gonze, X. and Vigneron, J.-P.},
  journal = {Phys. Rev. B},
  volume = {39},
  issue = {18},
  pages = {13120--13128},
  numpages = {0},
  year = {1989},
  month = {Jun},
  publisher = {American Physical Society},
  doi = {10.1103/PhysRevB.39.13120},
  url = {https://link.aps.org/doi/10.1103/PhysRevB.39.13120}
}

@article{PhysRevB.55.10355,
  title = {Dynamical matrices, Born effective charges, dielectric permittivity tensors, and interatomic force constants from density-functional perturbation theory},
  author = {Gonze, Xavier and Lee, Changyol},
  journal = {Phys. Rev. B},
  volume = {55},
  issue = {16},
  pages = {10355--10368},
  numpages = {0},
  year = {1997},
  month = {Apr},
  publisher = {American Physical Society},
  doi = {10.1103/PhysRevB.55.10355},
  url = {https://link.aps.org/doi/10.1103/PhysRevB.55.10355}
}

@article{bare,
  title = {Ab initio downfolding for electron-phonon-coupled systems: Constrained density-functional perturbation theory},
  author = {Nomura, Yusuke and Arita, Ryotaro},
  journal = {Phys. Rev. B},
  volume = {92},
  issue = {24},
  pages = {245108},
  numpages = {15},
  year = {2015},
  month = {Dec},
  publisher = {American Physical Society},
  doi = {10.1103/PhysRevB.92.245108},
  url = {https://link.aps.org/doi/10.1103/PhysRevB.92.245108}
}

@article{TaS2,
  title = {Ab initio phonon self-energies and fluctuation diagnostics of phonon anomalies: Lattice instabilities from Dirac pseudospin physics in transition metal dichalcogenides},
  author = {Berges, Jan and van Loon, Erik G. C. P. and Schobert, Arne and R\"osner, Malte and Wehling, Tim O.},
  journal = {Phys. Rev. B},
  volume = {101},
  issue = {15},
  pages = {155107},
  numpages = {14},
  year = {2020},
  month = {Apr},
  publisher = {American Physical Society},
  doi = {10.1103/PhysRevB.101.155107},
  url = {https://link.aps.org/doi/10.1103/PhysRevB.101.155107}
}

@misc{ToptoNot,
  author       = {Hu, Wenjie and Gong, Jiayi and Qiu, Yuhui and Yang, Lexian and Zhou, Jin-Jian and Yao, Yugui},
  title        = {Phonons Drive the Topological Phase Transition in Quasi-One-Dimensional Bi$_4$I$_4$},
  year         = {2025},
  howpublished = {arXiv},
  note         = {2025, arXiv:2508.17268. arXiv. \url{https://arxiv.org/abs/2508.17268} (accessed December 15, 2025).},
  url          = {https://arxiv.org/abs/2508.17268},
  urldate      = {2025-12-15}
}

\end{document}